\documentclass[aps, prx, preprint, onecolumn, nofootinbib]{revtex4-2}
\usepackage{amsmath,amssymb,bm,bbm} 
\usepackage{physics}
\usepackage{graphicx}  
\usepackage{float}
\usepackage{dsfont} 
\usepackage{color}
\usepackage{xcolor}
\usepackage[papersize={8.5in,11in}]{geometry}
\usepackage[colorlinks=true]{hyperref}  
\usepackage{comment}
\usepackage[section]{placeins}
\usepackage{tikz}

\usepackage{tikz-3dplot}
\usetikzlibrary{decorations.markings, decorations.pathreplacing}
\newcommand\hexsize{14pt}
\tdplotsetmaincoords{72}{132} 
\tdplotsetrotatedcoords{10}{0}{0} 
\hypersetup{
    unicode=false,          
    pdftoolbar=true,        
    pdfmenubar=true,        
    pdffitwindow=false,     
    pdfstartview={FitH},    
    pdfsubject={},   
    pdfcreator={},   
    pdfproducer={}, 
    pdfkeywords={} {} {}, 
    pdfnewwindow=true,      
    colorlinks=true,       
    linkcolor=magenta, 
    citecolor=blue,        
    filecolor=magenta,      
    urlcolor=blue           
} 

\usepackage{amsfonts}
\usepackage{upgreek}
\usepackage{slashed}
\usepackage{latexsym}

\newcommand{\beq}{\begin{equation}}
\newcommand{\eeq}{\end{equation}}
\def\bea{\begin{eqnarray}}
\def\eea{\end{eqnarray}}

\newcommand{\calD}{\mathcal{D}}
\newcommand{\UU}{\operatorname{U}}
\newcommand{\SU}{\operatorname{SU}}
\newcommand{\SO}{\operatorname{SO}}
\newcommand{\OO}{\operatorname{O}}

\newcommand{\USP}{\operatorname{USp}}

\definecolor{orange}{rgb}{1,0.5,0}

\begin{document}

\title{Sign-problem-free effective models of\texorpdfstring{\\}{} triangular lattice quantum antiferromagnets}

\author{Leyna Shackleton}
\affiliation{Department of Physics, Harvard University, Cambridge MA 02138, USA}

\author{Subir Sachdev}
\affiliation{Department of Physics, Harvard University, Cambridge MA 02138, USA}

\date{November 6, 2023}

\begin{abstract}
The triangular lattice antiferromagnet with $S=1/2$ spins and nearest neighbor interactions is known to have long-range antiferromagnetic order, with nearest-neighbor spins at an angle of 120 degrees. Numerical studies of quantum phases proximate to this state have been limited to small systems because the of the sign-problem in Monte Carlo simulations in imaginary time. We propose an effective lattice model for quantum fluctuations of the antiferromagnetic order, and a sign-problem free Monte Carlo algorithm, enabling studies in large systems sizes. The model is a $\mathbb{Z}_2$ gauge theory coupled to gauge-charged scalars which have a relativistic dispersion in the continuum limit. Crucially, the gauge theory is {\it odd} {\it i.e.\/} there is a static, background $\mathbb{Z}_2$ gauge charge on each site, accounting for the Berry phases of the half-odd-integer spins on each site. We present results of simulations on lattices of sizes up to $36 \times 36 \times 36$. Along with the antiferromagnetically ordered phase, our phase diagram has a valence bond solid state with a $\sqrt{12} \times \sqrt{12}$ unit cell, and a gapped $\mathbb{Z}_2$ spin liquid. Deconfined critical points or phases in intermediate regions are not ruled out by our present simulations.
\end{abstract}
\maketitle
\newpage
\section{Introduction}
\textit{Quantum spin liquids}~\cite{savary2016a,QPMbook} (QSLs) are exotic phases of matter which arise when strong frustration in a quantum spin system prevents the emergence of a conventional long-range ordered phase at zero temperature. Among the various platforms proposed to realize these unconventional phases, the geometric frustration present in triangular lattice Heisenberg antiferromagnets make them a natural candidate for QSL behavior. Experimental realizations in Yb-based compounds~\cite{li2015, shen2016, paddison2017, bordelon2019, ding2019, bordelon2020, dai2021, ranjith2019, ranjith2019a, zhang2021, zhang2022, scheie2023, xu2023} as well as organic compounds~\cite{shimizu2003, shimizu2006, itou2008, yamashita2009} have yielded promising results, including a lack of magnetic order and a continuum of low-energy spin excitations suggestive of fractionalized spinon degrees of freedom. Although the ground state of the spin $S=1/2$ Heisenberg antiferromagnet on the triangular lattice with only nearest-neighbor interactions is known to host conventional coplanar magnetic order~\cite{huse1988, bernu1994, capriotti1999}, the strength is reduced substantially by quantum fluctuations, and only a small amount of additional frustration from next-nearest-neighbor interactions is necessary to destroy the magnetic order~\cite{iqbal2016a, kaneko2014, zhu2015, hu2015, saadatmand2016, oitmaa2020, wietek2023}. The nature of the non-magnetic phase has been the source of much debate, both in these idealized lattice models and in aforementioned experimental realizations. In particular, there have been conflicting results on whether the fractionalized spinon excitations are gapped, have gapless Dirac nodes, or form a spinon Fermi surface.

Parton constructions provide a robust theoretical technique for describing a large class of phases of frustrated antiferromagnets, as well as phase transitions between them. The spin-$1 / 2$ operator $\vec{S}_i$ can formally be expressed as a bilinear operator in terms of either bosonic or fermionic spinons, with the constraint of one spinon per site enforced by the introduction of gauge fields. When the gauge field is deconfined, the system is a quantum spin liquid with fractionalized spin-$1 / 2$ spinon excitations. Various ordered phases, such as antiferromagnetism and valence bond solid (VBS) ordering, can be understood as instabilities to this deconfined phase.

Using this parton construction as a starting point, effective lattice models for describing the possible phases of quantum antiferromagnets can be deduced by minimally coupling bosonic spinons to emergent gauge fields. These effective models have the advantage of being more amenable to numerical simulations, as demonstrated in~\cite{park2002}, where an effective lattice model describing quantum antiferromagnets on the square lattice with easy-plane $\UU(1)$ symmetry was simulated numerically using Monte Carlo techniques. Large-scale simulations of the non-compact $\text{CP}^1$ model, conjectured to describe the deconfined quantum critical point separating N\'eel and VBS order on the square lattice, have also been studied~\cite{troyer2008, chen2013, nakane2009, bojesen2013} through Monte Carlo sampling. Outside the context of quantum magnetism, much progress has been made in developing numerical methods for simulating bosonic matter coupled to gauge fields~\cite{endres2007, delgadomercado2013}.

Following this approach, we present the results of a Monte Carlo simulation of an effective model of $\SU(2)$ antiferromagnetism on the triangular lattice. This effective model is derived using a bosonic spinon representation of the spin-$1/2$ degrees of freedom, where a mean-field analysis~\cite{sachdev1992} yields a gapped QSL phase with $\mathbb{Z}_2$ gauge fluctuations and an emergent $\OO(4)$ symmetry that rotates between the two low-energy bosonic spinon excitations. Our effective model which captures the QSL phase as well as ordered phases arising from either spinon condensation (coplanar magnetic order) or gauge confinement (non-magnetic VBS order) is that of an $\OO(4)$ vector field coupled to an \textit{odd} $\mathbb{Z}_2$ gauge field \cite{RJSS91,MVSS99,TSMPAF99,FHMOS04}. The odd nature of this $\mathbb{Z}_2$ gauge field is a consequence of the half-integer spin, and is essential in preventing the existence of a trivial disordered phase. The odd gauge field also leads to the appearance of a Berry phase in the action, which prohibits a direct Monte Carlo sampling of the partition function due to a sign problem. One of the primary contributions of this work is to present a sign-problem-free representation of this model, which is applicable to  

\begin{figure}
    \includegraphics[width=0.7\textwidth]{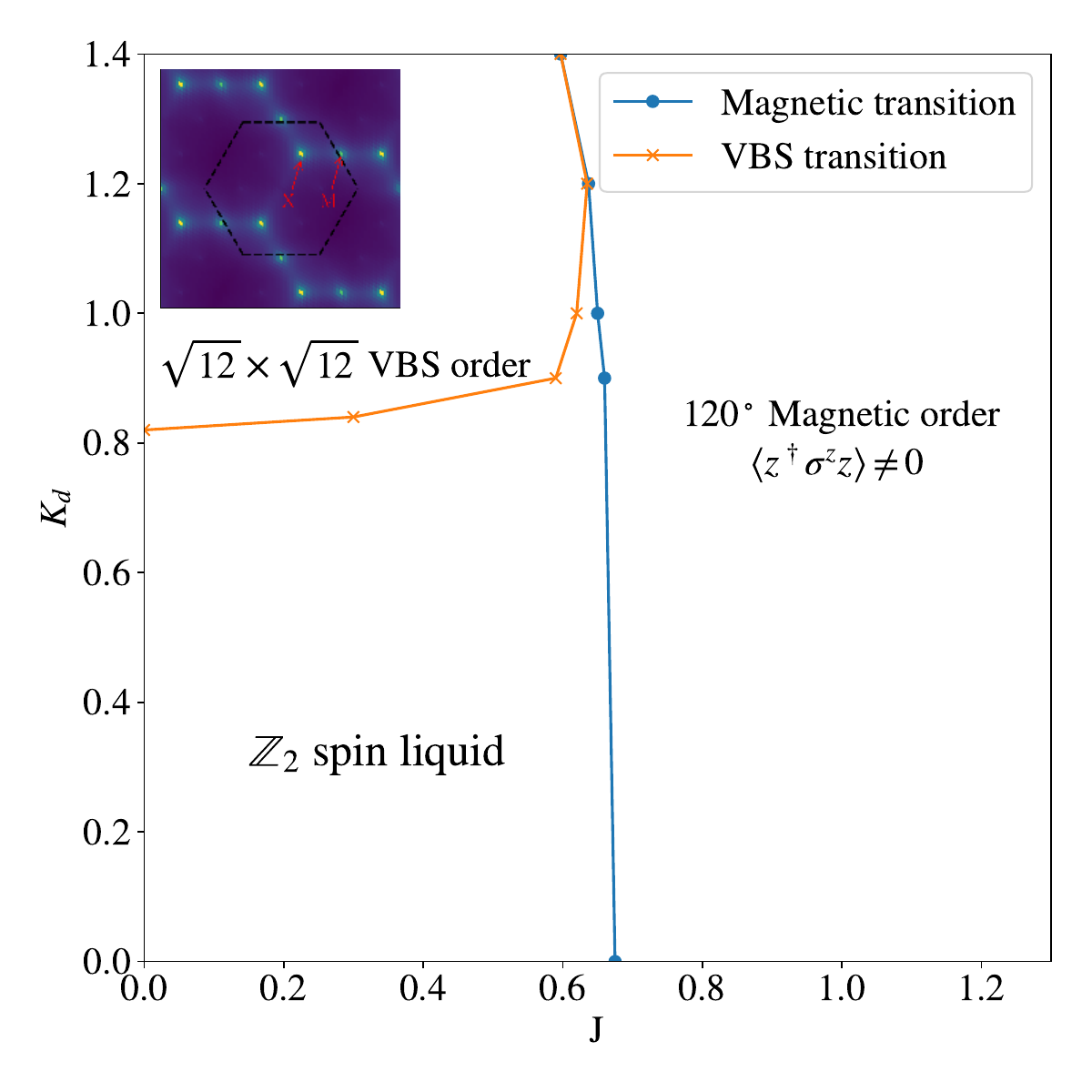}
    \caption{We plot the phase diagram of an $\OO(4)$ vector model coupled to an odd $\mathbb{Z}_2$ gauge field on a triangular lattice, as determined through classical Monte Carlo simulations. The model exhibits three phases, which correspond to a quantum spin liquid, $\sqrt{12} \times \sqrt{12}$ VBS, and coplanar antimagnetic order when regarded as an effective model of spin-$1 / 2$ Heisenberg antiferromagnetism on the triangular lattice. Algorithmic limitations discussed later prevent clear establishment of the location of the magnetic phase transition.}
    \label{fig:phaseDiagram}
\end{figure}

The primary result of this work - the phase diagram as a function of boson hopping $J$ and gauge action $K_d$ (to be precisely defined later) - is given in Fig.~\ref{fig:phaseDiagram}. All three phases - QSL, magnetic order, and VBS order - are present, with the valence bond ordering being of the $\sqrt{12} \times \sqrt{12}$ form, consistent with the pure gauge theory ($J = 0$) limit, which has been studied extensively~\cite{moessner2001, ralko2005, ralko2006, misguich2008, coletta2011}. Of note is a direct transition between the VBS and magnetic phases, which has been argued~\cite{jian2018, song2018, song2019, wietek2023} to be described by an emergent quantum electrodynamics with $N_f=4$ flavors of massless Dirac fermions. Our numerical results give some evidence for a first-order transition, although for reasons we will describe later, accurate Monte Carlo simulations of this model pose a number of challenges and we do not believe a continuous transition can be definitively ruled out.

The structure of this paper is as follows. In Section~\ref{sec:effectiveModel}, we make explicit the connection between our effective model and the microscopic spin degrees of freedom. In Section~\ref{sec:signProblem}, we describe the duality transformations that render the effective model sign-problem-free. Although these techniques are of general interest, we stress to the reader that the results of the Monte Carlo simulations, presented in~\ref{sec:numerics}, may be understood independent of the duality transformations.


\section{Connection between effective model and quantum magnetism}
\label{sec:effectiveModel}
We first outline a derivation of the effective model to be studied, and analyze its possible phases. Our starting point is the spin-$1 / 2$ Heisenberg antiferromagnet on the triangular lattice,
\begin{equation}
  \begin{aligned}
    H = \sum_{i j } J_{i j} \vec{S}_i \cdot \vec{S}_j\,,
    \label{eq:antiferro}
  \end{aligned}
\end{equation}
with $J_{i j}$ short-ranged antiferromagnetic interactions. The relation between the $\SU(2)$ triangular lattice antiferromagnet and a theory of bosonic spinons coupled to a $\mathbb{Z}_2$ gauge field was first derived in~\cite{sachdev1992} - by generalizing the $\SU(2)$ theory to $\USP(2M)$ and proceeding via a combined large-$M$ and large-$S$ expansion, the model becomes analytically tractable. For completeness, we provide a derivation of this in Appendix~\ref{sec:derivation}, and summarize the main points here:
\begin{itemize}
  \item The bosonic spinon representation introduces a dynamical $\UU(1)$ gauge field. The gapless photon excitations arising from these gauge fluctuations can present an obstacle for realizing a stable spin liquid phase. However, the saddle-point solutions for the gauge field - justified in a large-$M$, $S$ expansion - spontaneously break the $\UU(1)$ fluctuations down to $\mathbb{Z}_2$, where the gauge excitations (visons) are gapped.
  \item The bosonic spinon dispersion in the presence of this saddle-point solution has two minima at non-zero momenta. Writing an effective action in terms of these low-energy spinons, the lowest-order quartic interaction terms allowed by symmetry preserves an $\OO(4)$ symmetry that rotates between the two complex bosons, which contains both the $\SO(3)$ spin rotation and $120^\circ$ lattice rotation symmetry.
\end{itemize}

From these points, can write down an effective model for triangular lattice antiferromagnetism, which we will show can support magnetic order, VBS order, and spin liquids. This model consists of an $\OO(4)$ vector on each site of the triangular lattice, which parameterizes the bosonic spinon fluctuations. These degrees of freedom are minimally coupled to a $\mathbb{Z}_2$ gauge field. Importantly, this is an \textit{odd} $\mathbb{Z}_2$ gauge field, which arises from a background spinon density of one spinon per site. The odd nature of this gauge theory prevents the confining phase from being a trivial gapped phase, in agreement with LSM theorems that prohibits such a phase for half-integer spins.

This model supports three types of phases. When the $\OO(4)$ spinons are uncondensed and the gauge field is deconfined, the system is a gapped $\mathbb{Z}_2$ spin liquid with topological order. Condensing the bosons spontaneously breaks the $\OO(4)$ symmetry, which in turn breaks both the $\SO(3)$ spin rotation symmetry and $120^\circ$ lattice rotation symmetry. Because of the $\mathbb{Z}_2$ gauge redundancy, the ground state manifold (GSM) for this order is $S^3 / \mathbb{Z}_2 = \SO(3)$ - in agreement with the $\SO(3)$ GSM of the $120^{\circ}$ magnetic order. The confining phase of the $\mathbb{Z}_2$ gauge field preserves spin rotation symmetry, but due to the odd nature, breaks lattice symmetries rather than being trivial - it is a valence bond solid phase. The pattern of lattice symmetry breaking is known in the pure gauge theory limit with nearest-neighbor interactions to be a $\sqrt{12} \times \sqrt{12}$ order~\cite{moessner2001}, with a 12-site unit cell, although effective longer-range interactions generated by the spinons can lead to different symmetry breaking patterns~\cite{slagle2014}.

We study this model using Monte Carlo techniques. The partition function for this two-dimensional quantum model on the triangular lattice can be mapped to an equivalent classical model on a three-dimensional stacked triangular lattice,
\begin{equation}
  \begin{aligned}
    \mathcal{Z} &= \sum_{s_{j, j + \hat{\mu}} = \pm 1} \prod_j \int \dd{z}_{j \alpha}\delta \left( \sum_\alpha \abs{z_{j \alpha}^2} - 1 \right) \left[ \prod_j s_{j, j + \tau} \right] \exp\left( - H[z_\alpha, s] \right)  \\
    H[z_\alpha, s] &= - \frac{J}{2} \sum_{ \langle j , \mu \rangle} s_{j, j + \hat{\mu}} \left( z_{j,  \alpha}^* z_{j + \hat{\mu},  \alpha} + \text{ c.c} \right) - K \sum_{\triangle \square} \prod_{\triangle \square} s_{j, j + \hat{\mu}} \,,
    \label{eq:bosonZ}
  \end{aligned}
\end{equation}
The two bosonic spinons $z_{j\alpha}$, $\alpha = 1\,, 2$ are minimally coupled to a classical $\mathbb{Z}_2$ gauge field $s_{j, j + \hat{\mu}}$ living on the links of the stacked triangular lattice. The odd nature of the gauge field is captured by the Berry phase term $\prod_j s_{j, j + \tau}$. Note that this Berry phase takes on values $\pm 1$, and hence the partition function as written in Eq.~\ref{eq:bosonZ} is not amenable to Monte Carlo simulations. As such, we must perform a series of transformations to obtain a sign-problem-free representation, where the partition function is expressed as a sum over purely positive weights.

\section{Sign-problem-free mapping}
\label{sec:signProblem}
In this section, we describe the mapping from the partition function in the previous section to one consisting only of positive weights. This is a very general mapping, valid for any $\text{O}(2n)$ vector model with integer $n$ on a large class of lattices, including the stacked triangular lattice relevant to our study of quantum antiferromagnetism on the 2D triangular lattice. However, the general approach does not differ substantially from the simplest case, which is an $\text{O}(2)$ model on the 3D cubic lattice. As the notation required to state the mapping in its most general form is rather complex, we find it most clear to first describe the sign-free mapping of an $\text{O}(2)$ model coupled to an odd $\mathbb{Z}_2$ gauge field on a 3D cubic lattice, and then subsequently describe the modifications necessary for alternate lattices or for general $\text{O}(2n)$ models. The mapping in this simpler limit was first carried out in~\cite{sachdev2002, park2002}, but using a different approach that does not as easily generalize to $\OO(2n)$ models. We outline a more generalizable mapping which also more carefully treats subtleties involving periodic boundary conditions. The initial steps of this mapping follow along the same lines as well-known particle-vortex dualities~\cite{dasgupta1981, jose1977, peskin1978}, which map an $\OO(2)$ model to a dual $\OO(2)$ model coupled to an emergent $\UU(1)$ gauge field. In this language, our $\mathbb{Z}_2$ gauge field couples to the emergent $\UU(1)$ gauge field via a mutual Chern-Simons term - we demonstrate that this allows for the $\mathbb{Z}_2$ gauge field to be integrated out, yielding a sign-problem-free representation.
\subsection{\texorpdfstring{$\text{O}(2)$}{O(2)} model}
Our model is described by the partition function
\begin{eqnarray}
&& {\cal Z} = \sum_{\{ s_{j,j+\hat{\mu}} = \pm 1\}} \int \prod_j d
\theta_j \exp \Biggl( K \sum_{\square} \prod_{\square}
s_{j,j+\hat{\mu}} \nonumber
\\ &&\!\!\!\!\!\!\!\!\!
+ \frac{4}{g} \sum_{j, \hat{\mu}} s_{j,j+\hat{\mu}} \cos \left(
\frac{\Delta_{\mu} \theta_j}{2} \right) - i \frac{\pi}{2} \sum_j (
1 - s_{j,j+\hat{\tau}}) \Biggr), \label{u1}
\end{eqnarray}
with, as in the previous section, $\Delta_\mu$ denoting the discrete lattice derivative, and $\prod_\square$ denoting the product of spins around a plaquette. This is an XY model whose degrees of freedom are angular variables $\theta_j$, coupled to an odd $\mathbb{Z}_2$ gauge field. The final term, corresponding to the Berry phase of the background boson filling, gives negative weights to the summation, thereby preventing sampling through Monte Carlo techniques.

Our first step is to rewrite the action for $\theta_j$ using the identity
\begin{equation}
    e^{s A \cos\theta} \propto \sum_{p = -\infty}^\infty e^{i p  \left( \theta + \pi\frac{1 - s}{2}\right)} I_p(A)
    \label{eq:besselIdentity}
\end{equation}
where $A > 0$, $s = \pm 1$, and $I_p(A)$ is the modified Bessel function of the first kind. The asymptotic behavior of $I_p(A)$ as $A \rightarrow \infty$ contains the more standard action for $p$ when the Villain approximation is used,
\begin{equation}
    I_p(A \gg 1) \propto \exp  \left[ - \frac{p^2}{2A}\right]\,.
\end{equation}
For the $\text{O}(2)$ model, this approximation does not alter the phase diagram, so we will use this for notational simplicity and to connect to prior work. However, the following mapping can be carried out while keeping the full Bessel function explicit, which will be necessary for general $\text{O}(2n)$ models as an analogous approximation breaks the $\text{O}(2n)$ symmetry down to $\text{O}(2)^{\otimes n}$.

We apply this identity to each instance of $s_{j, j + \hat{\mu}} \cos \left( \frac{\Delta_\mu \theta_j}{2} \right)$, thereby introducing an integer-valued field $p_\mu$ on each of the links. This allows $\theta_j$ to be integrated out at the cost of imposing a divergence-free constraint $\Delta_\mu p_\mu = 0$. This divergence-free constraint implies that $p_\mu$ must form closed integer-valued ``current loops.'' This requirement can be made explicit by writing $p_\mu$ as the curl of a height field, $h_{\overline{j}\mu}$ living on the links of the dual lattice. This allows for the creation of local current loops - for periodic boundary conditions, one must also include the possibility of non-contractible loops that cannot be expressed as the curl of a height field. Assuming we have periodic boundary conditions in all three directions, it is sufficient to pick a representative non-contractible current loop in each direction, $(w^x_{j \mu}, w_{j \mu}^y, w_{j \mu}^\tau)$ and express
\begin{equation}
    p_{j \mu} \equiv \epsilon_{\mu\nu\lambda} \Delta_\nu h_{\overline{j} \lambda} + \vec{n} \cdot \vec{w}_{j \mu}
    \label{eq:currentConfig}
\end{equation}
where $\vec{n} \equiv (n_x\,, n_y\,, n_\tau)$ is an integer-valued vector specifying the winding number of the current in each of the three dimensions. This representation introduces additional degrees of freedom, as one may always shift $h_{\overline{j} \mu}$ by the divergence of a scalar field without changing the current configuration. In the language of particle-vortex dualities, this reflects the emergent $\UU(1)$ gauge redundancy.
These additional degrees of freedom are crucial to our sign-free mapping.

Our partition function at this point is
\begin{eqnarray}
&& {\cal Z} = \sum_{\{ s_{j,j+\hat{\mu}} = \pm 1\}} \sum_{h_{\overline{j} \mu} = - \infty}^\infty \sum_{n_i = -\infty}^\infty \exp \Biggl( K \sum_{\square} \prod_{\square}
s_{j,j+\hat{\mu}} \nonumber
\\ &&\!\!\!\!\!\!\!\!\!
- \frac{g}{8} \sum_{j, \hat{\mu}} p_{j \mu}^2 + i \frac{\pi}{2} \sum_{j, \hat{\mu}} p_{j \mu} (1 - s_{j, j + \hat{\mu}})- i \frac{\pi}{2} \sum_j (
1 - s_{j,j+\hat{\tau}}) \Biggr).
\end{eqnarray}
In this form, the Berry phase term can be absorbed by a shift $h_{\overline{j} \mu} \rightarrow h_{\overline{j} \mu} + h^0_{\overline{j} \mu}$, where $\epsilon_{\mu\nu\lambda} \Delta_\nu h^0_{\overline{j} \lambda} = \delta_{\mu, \tau}$.

We now address the coupling between the current $p_{j \mu}$ and the gauge field, which corresponds to a mutual Chern-Simons term in the vortex representation. For the local current loops expressible in terms of the height field, we use the identity
\begin{equation}
  \exp[ i \pi \sum_{j, \hat{\mu}}\epsilon_{\mu\nu\lambda} \Delta_\nu h_{\overline{j} \lambda} \frac{1 - s_{j, j + \hat{\mu}}}{2}] =
 \exp\left[ i \pi \sum_{j \hat{\mu}} h_{\overline{j}\mu} \frac{1 - \prod_\square s_{j, j + \hat{\mu}}}{2} \right]\,,
 \label{eq:curlRewriting}
\end{equation}
which can be verified by expanding out the curl and collecting terms proportional to $h_{\overline{j} \mu}$. To address the coupling to non-contractible current loops, we first note that after performing the transformation in Eq.~\ref{eq:curlRewriting}, the coupling to the non-contractible current loops is the only place where the gauge field appears explicitly - all other terms in the partition function only depend on the plaquette flux $\prod_\square s_{j, j + \hat{\mu}}$. We argue that if any component of $\vec{n}$ is odd, the contribution to the partition function vanishes. This is because for an arbitrary configuration and for a choice of direction $\hat{\nu}$, one can make a ``large'' gauge transformation consisting of flipping all $s_{j, j + \hat{\nu}}$ spins that intersect a plane tangent to $\hat{\nu}$. This keeps all the plaquette fluxes invariant but flips the sign of a single $s_{j, j + \hat{\nu}}$ that intersects the the non-contractible current loop in the $\hat{\nu}$ direction. If the winding number in this direction is odd, the new configuration contributes to the partition function with the same magnitude but opposite sign as the original one, leading to an exact cancellation. We therefore restrict our sum over $n_i$ to be even, in which case the coupling to the gauge field drops out entirely as it always contributes a factor of $1$. If one interprets the current loops as bosonic worldlines, this constraint is simply the statement that bosons must be created in pairs and sectors with odd numbers of bosons are unphysical. 

We now integrate out the gauge field. Note that only the plaquette flux $\frac{1 - \prod_\square s_{j, j + \hat{\mu}}}{2}\equiv \Phi_{\overline{j}, \mu}$ appears in the partition function (the dual link $\overline{j}, \mu$ uniquely labels a plaquette). Although normally incorrect, we claim that it is valid to perform an independent summation over all possible plaquette flux values $\Phi_{\overline{j}, \mu} = 0\,,1$ on each plaquette. This is not true in general, as there is a non-trivial constraint on the possible values of flux - starting from the flux-free configuration with $s_{j, j + \hat{\mu}} = 1$, $\Phi_{\overline{j}, \mu} = 0$, gauge fluctuations can only change the divergence of $\Phi$ at any dual site by multiples of two, i.e.,
\begin{equation}
\nabla_\mu \Phi_{\overline{j}, \mu} = 0 \text{ mod } 2\,.
\label{eq:fluxConstraint}
\end{equation}

The key observation is that the redundant degrees of freedom introduced in the height field representation for $p_{j, \mu}$ serve as Lagrange multipliers to dynamically enforce Eq.~\ref{eq:fluxConstraint}. As a consequence, one can directly perform the summation over all gauge field configurations. To see how this constraint is enforced, let us make the redundant degrees of freedom explicit by writing $h_{\overline{j}, \mu} = \tilde{h}_{\overline{j}, \mu} + \Delta_\mu f_{\overline{j}}$, where we perform the summation over both distinct current loop configurations $\tilde{h}$ and the redundant degrees of freedom $f$. The coupling of $f$ to the gauge field is
\begin{equation}
    \exp\left[ i \pi \sum_{j \mu} \Delta_\mu f_{\overline{j}}  \Phi_{\overline{j}, \mu} \right] = \exp\left[  - i \pi \sum_{j \mu} f_{\overline{j}} \Delta_\mu \Phi_{\overline{j} \mu} \right] \,.
\end{equation}
Performing a summation over $f_{\overline{j}}$ will impose the constraint Eq.~\ref{eq:fluxConstraint}. 

Integrating out the gauge field gives the final sign-free representation of our partition function,
\begin{equation}
\begin{aligned}
    \mathcal{Z} &= \sum_{h_{\overline{j} \mu} = - \infty}^\infty \sum_{n_i = -\infty}^\infty \exp \Bigg( -\frac{g}{8} \sum_{j, \mu} p_{j \mu}^2 + K_d \sum_{j, \mu} \varepsilon_{\overline{j}, \overline{j} + \mu} \sigma_{\overline{j}, \overline{j} + \mu} \Bigg)
    \end{aligned}
\end{equation}
where
\begin{equation}
\begin{aligned}
    \tanh K_d &= e^{-2 K}\,,
    \\
    \sigma_{\overline{j}, \overline{j} + \mu} &= 1 - 2 (h_{\overline{j} \mu} \text{ mod } 2)\,,
\end{aligned}
\end{equation}
$p_{j\mu}$ is defined in terms of $h_{\overline{j} \mu}$ and $\vec{n}$ as in Eq.~\ref{eq:currentConfig}, and $\varepsilon$ is a \textit{fixed} field taking values $\pm 1$, with the constraint that the product of $\varepsilon$ around any temporal (spatial) plaquette is +1 (-1). The factor of $\varepsilon$ arises because of the background height field $h^0$, which in turn is a consequence of the Berry phase. Because of this, the model in the limit $g \rightarrow \infty$ (i.e. when the $\text{O}(2)$ coupling drops out and we recover a pure $\mathbb{Z}_2$ gauge field) reduces down to a 3D Ising model with frustration on the xy-planes, which is dual to an odd $\mathbb{Z}_2$ gauge theory, rather than a frustration-free Ising model dual to an even $\mathbb{Z}_2$ gauge theory.
\begin{figure}
    \centering
    \includegraphics[width=0.4\textwidth]{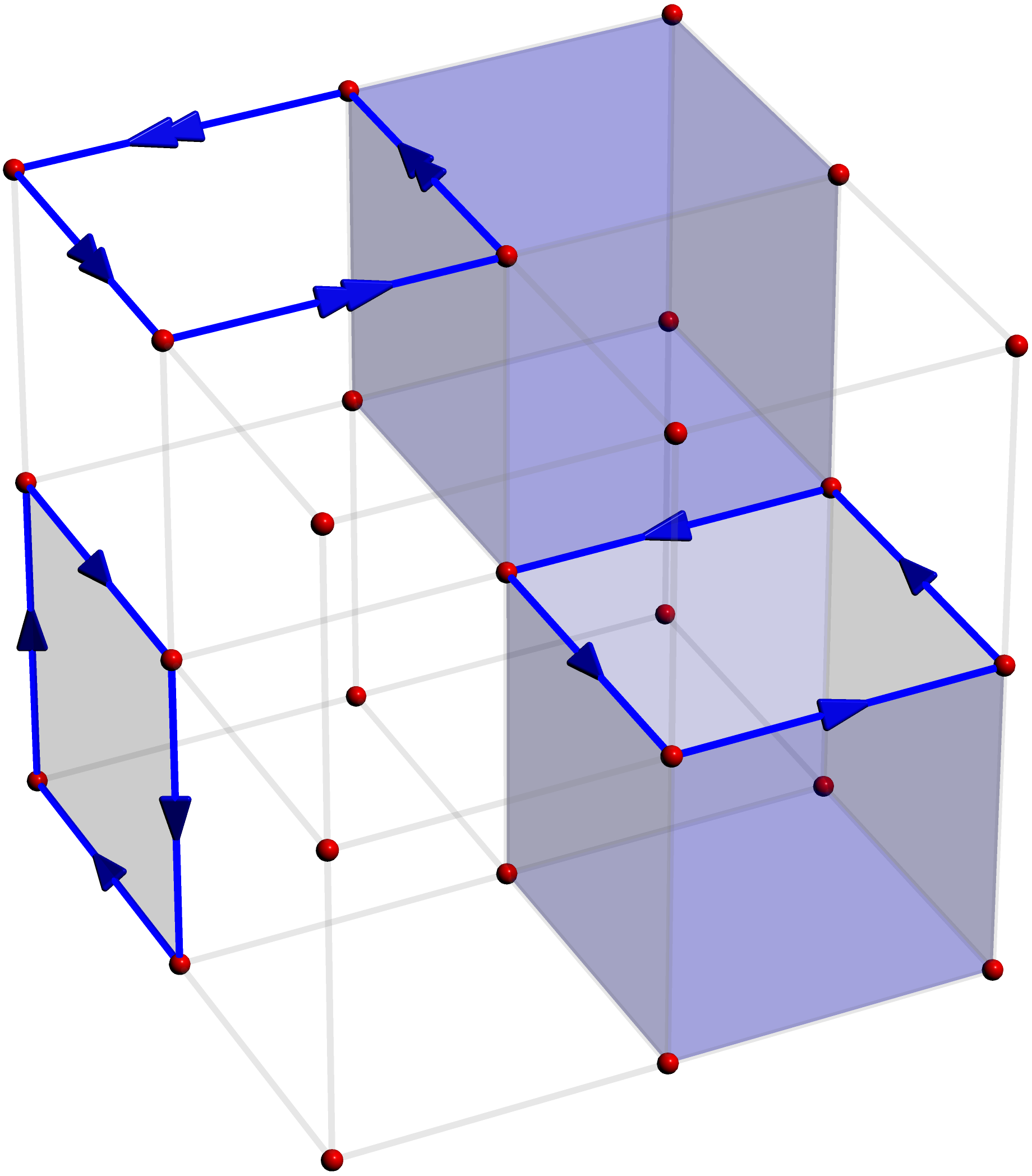}
    \hspace{30pt}
        \includegraphics[width=0.4\textwidth]{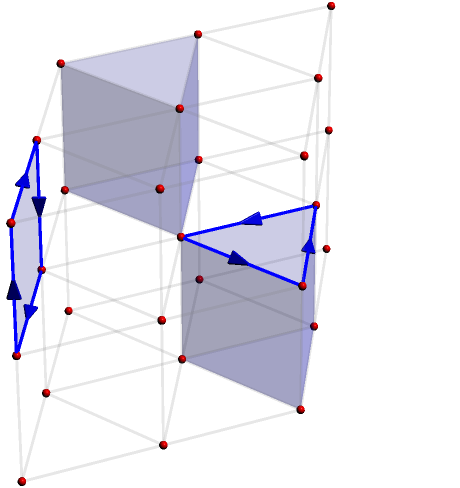}
    \caption{The dual representation of an $\OO(2)$ model coupled to $\mathbb{Z}_2$ gauge field can be expressed in terms of integer-valued current loops and $\mathbb{Z}_2$ membranes of flux, with the constraint that odd-valued current loops must form the boundary of an open $\mathbb{Z}_2$ surface. The effect of a Berry phase is to introduce frustration in the surface action. This dual mapping is suitable for both the square lattice (left) as well as the triangular lattice (right). }
    \label{squareLattice}
\end{figure}

This theory has a simple interpretation, illustrated in Fig.~\ref{squareLattice}. One can interpret the field $\sigma_{\overline{j}, \overline{j} + \mu}$ as a $\mathbb{Z}_2$ flux variable living on the plaquettes of the lattice, and the relationship between $p$ and $\sigma$ translates into the constraint that any odd current loop must form the perimeter of an open ``surface'' of flux. Closed surfaces correspond to height field configurations with vanishing curl. For an \textit{even} $\mathbb{Z}_2$ gauge theory, the coupling constant $K_d$ has the simple interpretation of a surface tension, with an energy cost proportional to the surface area. When only closed surfaces are allowed, these surfaces also have the interpretation of domain walls of the dual Ising model. For an odd $\mathbb{Z}_2$ gauge theory, the background field $\varepsilon$ introduces frustration - plaquette fluxes cost positive or negative energy depending on the location, and it is no longer possible to energetically satisfy all plaquettes using only closed surfaces.
\subsection{Generalization to \texorpdfstring{$\text{O}(2n)$}{O(2n)} models}
We now show how this sign-free mapping lifts to a model where our degrees of freedom consist of $\OO(2n)$ spins, with $n \geq 1$. While particle-vortex dualities do not have a generalization to non-Abelian $\OO(2n)$ models, one may think of this mapping as a sort of particle-vortex duality applied to an Abelian $\OO(2)^{\otimes n}$ subgroup. The full non-Abelian $\OO(2n)$ symmetry is preserved in virtue of working with an explicit lattice action and restricting oneself to exact transformations that necessarily keep the partition function invariant.

An $n$-component complex vector $z_{j \alpha}$, $1 \leq \alpha \leq n$, lives on each site, with the constraint that $\sum_\alpha \abs{z_{j\alpha}}^2 = 1$. Our partition function is
\begin{eqnarray}
&& {\cal Z} = \sum_{\{ s_{j,j+\hat{\mu}} = \pm 1\}} \int \prod_{j\,, \alpha} \dd{z_{j\alpha}} \delta(\sum_\alpha \abs{z_{j \alpha}}^2 - 1)\exp \Biggl( K \sum_{\square} \prod_{\square}
s_{j,j+\hat{\mu}} \nonumber
\\ &&\!\!\!\!\!\!\!\!\!
+ J\sum_{j, \hat{\mu}, \alpha} s_{j,j+\hat{\mu}} \left(z_{j \alpha}^* z_{j + \mu, \alpha} + \text{ c.c}\right) - i \frac{\pi}{2} \sum_j (
1 - s_{j,j+\hat{\tau}}) \Biggr), \label{o2n}
\end{eqnarray}
Our starting point is a representation $z_j$ that makes connection to our previous mapping, $z_{j\alpha} \equiv r_{j \alpha} e^{i \theta_{j \alpha}}$. Note that in this represntation, the magnitude fields $r_{j\alpha}$ are gauge-neutral, and only the phase variables $\theta_{j\alpha}$ are affected by gauge transformations. In terms of these variables,
\begin{eqnarray}
&& {\cal Z} = \sum_{\{ s_{j,j+\hat{\mu}} = \pm 1\}} \prod_{j, \alpha}\int_0^{2\pi} \dd{\theta_{j \alpha}}  \int_0^1  r_{j \alpha}\dd{r_{j \alpha}}\delta(\sum_\alpha \abs{r_{j \alpha}}^2 - 1)\exp \Biggl( K \sum_{\square} \prod_{\square}
s_{j,j+\hat{\mu}} \nonumber
\\ &&\!\!\!\!\!\!\!\!\!
+ J \sum_{j, \hat{\mu}, \alpha} s_{j,j+\hat{\mu}} r_{j, \alpha} r_{j + \mu, \alpha} \cos(\Delta_\mu \theta_{j \alpha})- i \frac{\pi}{2} \sum_j (
1 - s_{j,j+\hat{\tau}}) \Biggr), \label{o2nangular}
\end{eqnarray}
As before, we use the identity Eq.~\ref{eq:besselIdentity}, except we introduce $n$ different integer-valued fields $p_{j \mu \alpha}$ on each link. Making the Villain approximation is no longer appropriate in this case - doing so breaks the full $\text{O}(2n)$ symmetry down to $n$ copies of $\text{O}(2)$, as the coefficient in front of the cosine term is no longer just a coupling constant but rather the dynamical field $r_{j, \alpha} r_{j + \mu, \alpha}$. As such, we keep the Bessel functions $I_{p_{j, \alpha, \mu}}(J r_{j, \alpha} r_{j + \mu, \alpha})$ explicit in our rewriting. Such an approach has previously been used to study $\OO(2n)$ models at finite density~\cite{endres2007} - this current loop representation also serves to cure the sign problem present when one introduces a non-zero chemical potential. This representation obfuscates the full $\OO(2n)$ symmetry, but retains an $S_n$ subset coming from permutations of the $\alpha$ indices. 

The rest of the mapping proceeds in a similar manner. We introduce $n$ height field representations $h_{\overline{j} \mu \alpha}$, with the plaquette flux $\Phi_{\overline{J} \mu}$ coupling to the \textit{total} height field $\sum_{\alpha} h_{\overline{j} \mu \alpha}$. The Berry phase term can be absorbed by a shift in any of the $n$ height fields - the choice is arbitrary and does not change the final representation.

Integrating out the gauge field gives the final form of our partition function,
\begin{equation}
  \begin{aligned}
    \mathcal{Z} &= \sum_{h_{\overline{j}, \alpha, \mu} = - \infty}^\infty \prod_{j \alpha}  \int_0^1 r_{j, \alpha} \dd{r_{j, \alpha}}\delta \left( \sum_\alpha r_{j, \alpha}^2 - 1 \right)  \exp\left( - H[r_\alpha\,, h_\alpha] \right)  \\
  H[r_\alpha\,, h_\alpha] &= \sum_{ \langle j , \mu \rangle} \left[ -\ln I_{p_{j, \alpha, \mu}}(J  r_{j ,\alpha} r_{j + \hat{\mu}, \alpha}) + K_d \varepsilon_{\overline{j}, \mu} \sigma_{\overline{j}, \mu}  \right]   \,.
    \label{eq:signProblemFreeH}
  \end{aligned}
\end{equation}
where
\begin{equation}
  \begin{aligned}
    \sigma_{\overline{j}, \mu} &=1 -  2 \left(\sum_\alpha h_{\overline{j}, \alpha, \mu} \right)\text{ mod } 2 
    \\
    e^{-2 K_d} &=  \tanh K  \\
  \end{aligned}
\end{equation}
and $\varepsilon$ is a static field taking values $\pm 1$, such that the product of $\varepsilon$ around each spatial (temporal) dual plaquette is $-1$ ($+1$). 
\subsection{Generalization to alternate geometries}
\begin{figure}[ht] \centering

      \begin{tikzpicture}[
  x=\hexsize,
  y=\hexsize,
  square/.style={draw,fill,minimum size=1.5pt,inner sep=0pt},
  dot/.style={circle,circle,fill,minimum size=2pt,inner sep=0pt}
  ]
  \foreach \x in {0,...,4} {
    \coordinate(X) at (0:1.5*\x);
    \ifodd\x
      \def\ymax{4}
      \coordinate(X) at ($(X)+(0:0.5)+(-120:1)$);
    \else
      \def\ymax{3}
    \fi
    \foreach \y in {0,...,\ymax} {
      \coordinate (\x-\y) at ($(X)+(60:\y)+(120:\y)$);
      \draw[gray] (\x-\y) +(-60:1)
      \foreach \z [remember=\z as \lastz (initially 4)] in {0,...,4} {
        -- coordinate(\x-\y-\lastz-m) +(\z*60:1) coordinate(\x-\y-\z)
      } -- cycle;
    }
  }
  \foreach \y in {0,...,3} {
    \draw[ultra thick, color=red] (4-\y-0) -- (4-\y-1);
    \draw[ultra thick, color=red] (2-\y-0) -- (2-\y-1);
    \draw[ultra thick, color=red] (0-\y-0) -- (0-\y-1);
  }
    \draw[ultra thick, color=red] (1-0-3) -- (1-0-4);
    \draw[ultra thick, color=red] (3-0-3) -- (3-0-4);
\end{tikzpicture}
\caption{The presence of a Berry phase gives rise to frustration in the effective Ising model on the dual lattice. For our theory defined on a triangular lattice, we show one possible tiling of the dual (hexagonal lattice) lattice, where the field $\varepsilon_{\overline{j}, \overline{j} + \mu} = - 1$ on red links, and 1 otherwise. This is chosen so that the product around any spatial plaquette gives $-1$.}
  \label{fig:triangular}
\end{figure}
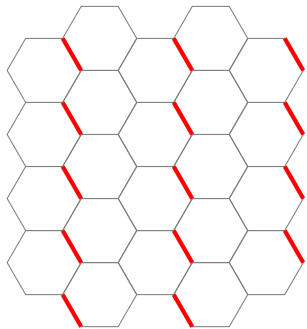

In our previous sections, we described this sign-free mapping on a cubic lattice, primarily for the simplicity in notation that the lattice provides. However, we emphasize that this mapping holds for more general lattices, including the stacked triangular lattice relevant to our current interest, as well as a stacked kagom\'e lattice. This mapping is easiest on lattices with even coordination, where integrating out the $\theta$ fields yields a familiar divergence-free constraint on $p$. In Fig.~\ref{fig:triangular}, we show this for a stacked triangular lattice, where the constraint is $\Delta_\mu p_{j \mu} = 0$, $\mu = \hat{e}_1\,, \hat{e}_2\,, \hat{e}_3 \,, \hat{\tau}$. and can again be satisfied by a height field representation. These height fields couple to the $\mathbb{Z}_2$ gauge flux in an identical manner, and inclusion of the Berry phase term manifests itself as a constant field $\varepsilon_{\overline{j}, \overline{j} + \mu}$ living on the dual lattice (stacked hexagonal lattice), with the constraint that the product around any spatial (temporal) plaquette is $-1$ ($+1$). An example configuration is shown in Fig.~\ref{fig:triangular}. While the dual lattice has odd coordination and a dual bond cannot technically be specified by the indices $(\overline{j}, \mu)$, we will continue to use the notation $h_{\overline{j}\mu}$ for simplicity as this subtlety will not be relevant.

\tdplotsetrotatedcoords{0}{0}{0} 
The lack of a bipartite lattice raises an important point in our duality mapping. Because our original lattice is not bipartite, the dual lattice does not have a natural definition of divergence - this can be seen from the stacked hexagonal lattice, dual to the triangular lattice, which has odd coordination so no symmetric definition of ``ingoing'' and ``outgoing'' bonds can be made. As a consequence, one must make sense of the use of dual lattice divergences in our derivation, which is employed in Eq.~\ref{eq:fluxConstraint}. Important to this is the $\mathbb{Z}_2$ nature of our gauge field - as a consequence, all divergences appear in equations that are only sensitive to whether the resulting expression is even or odd. Hence, the sign structure of the divergence operator on the dual lattice is irrelevant, as a different sign structure only changes the divergence by an even amount.

We may also verify our method of integrating out the $\mathbb{Z}_2$ gauge field without making any reference to a divergence. Our ``naive'' procedure of integrating out the $\mathbb{Z}_2$ gauge field by independently summing over all possible plaquette flux values $\Phi_{\overline{j}, \mu} = 0\,, 1$ is valid so long as unphysical flux configurations are dynamically cancelled out by the redundant degrees of freedom introduced by introducing a height field representation. The key feature we need is some notion of a ``gauge'' transformation $h_{\overline{j} \mu} \rightarrow h_{\overline{j} \mu} + \Delta_\mu f_{\overline{j}}$ that can leave the $p_{j \mu}$ current configuration invariant but change the parity of $h_{\overline{j} \mu}$ along an arbitrary closed surface. Unphysical flux configurations correspond to having an odd total flux along some closed surface, and for these configurations, performing a gauge transformation flips the sign of its contribution to the partition function and hence cancels out the unphysical configuration. For a triangular lattice, defining such a transformation in terms of the divergence of a scalar field on the dual lattice is not straightforward - however, one can easily verify that such a transformation is still possible by creating a current loop on each plaquette of the closed surface, with the orientations chosen in such away that all currents cancel out. The smallest such surface involves five plaquettes, and as such, there is no symmetric way of distributing the shifts $h_{\overline{j} \mu} \rightarrow h_{\overline{j} \mu} \pm 1$. Crucially, the $\mathbb{Z}_2$ gauge field only sees the \textit{parity} of $h_{\overline{j} \mu}$, so this subtlety is not an issue.

For lattices with odd coordination, such as the hexagonal lattice, one must take extra care with dealing with the divergence-free constraint on the original lattice. While we believe that our duality mapping will likely be applicable to these models as well, we defer a more thorough analysis for future work.

\section{Results from Monte Carlo simulations}
\label{sec:numerics}
We now present numerical results of the simulation of an $\OO(4)$ vector model coupled to an odd $\mathbb{Z}_2$ gauge field on the triangular lattice. The Hamiltonian is defined in Eq.~\ref{eq:signProblemFreeH} for two species of current loops.
This model has several simple limits:
\begin{itemize}
    \item $K_d = 0$: in this limit, the gauge field becomes static and our model reduces to that of an $\OO(4)$ model, albeit in an unconventional current loop representation. The presence of a $\mathbb{Z}_2$ gauge field still has the effect of restricting our observables to be gauge-invariant, and hence the critical theory for the boson condensation is given by  $\OO(4)^*$, which possesses the same critical exponents as the $\OO(4)$ universality class but for which differences can be found in terms of the excitation spectrum for a finite-size system~\cite{whitsitt2016}. This difference is reflected in our dual theory by the topological constraint that the winding number of current loops must be even.
    \item $K_d = \infty$: in this limit, confinement of the gauge field prevents individual bosonic excitations. The relevant degrees of freedom are the gauge-invariant $\SO(3)$ order parameters $z_i^\dagger \vec{\sigma} z_i$. For an even $\mathbb{Z}_2$ gauge field, this limit would be described by a non-linear $\sigma$ model, and would support an ordered and disordered phases. This trivial disordered phase is ruled out in our case from the LSM theorem, and the simple non-linear $\sigma$ model picture is modified for an odd $\mathbb{Z}_2$ gauge field by the influence of a Berry phase on vortices of the $\SO(3)$ order parameter. We expect that this limit will always give an ordered phase, which we will verify in future numerical studies.
  \item $J = 0$: in this limit, we expect to recover an odd $\mathbb{Z}_2$ gauge theory. In our dual formulation, current loop excitations cost infinite energy and our state space is restricted to configurations with $p_{j \mu \alpha} = 0$. Within this space, we have a single $\mathbb{Z}_2$ degree of freedom residing on each dual site $(\overline{j})$, the flipping of which at site $\overline{k}$ corresponds to the shift $h_{\overline{j} \mu \alpha} \rightarrow h_{\overline{j} \mu \alpha} + \Delta_{\mu} f_{\overline{j} \alpha}$, with $f_{\overline{j} \alpha} = \delta_{\overline{j} \overline{k}}$ (performed on a random choice of $\alpha$). Since the action is only sensitive to the parity of $\sum_{\alpha} h_{\overline{j} \mu \alpha}$, this is effectively a $\mathbb{Z}_2$ degree of freedom. Our model then reduces down to a frustrated Ising model defined on the dual lattice - dual to the odd $\mathbb{Z}_2$ gauge theory - which displays a transition at a critical value of $K_d$ from a disordered to an ordered phase. In terms of the original spin degrees of freedom, this is a transition from the gapped $\mathbb{Z}_2$ spin liquid to a VBS phase. Semiclassical analyses~\cite{moessner2001} predict that this transition is in the $\OO(4)$ universality class, and this prediction is supported by quantum dimer model simulations~\cite{yan2021}.
\end{itemize}

We analyze the Hamiltonian in Eq.~\ref{eq:signProblemFreeH}, for $n=2$ and defined on an $L \times L \times L$ stacked triangular lattice, by sampling configurations $\{r_{j \alpha}\,, h_{\overline{j} \mu \alpha}\}$. Simulations are done for $L=12$, $24$, and $36$ - keeping the linear system size a multiple of $12$ is necessary to accommodate the large unit cell of the $\sqrt{12} \times \sqrt{12}$ VBS order. Movement through the configuration space is accomplished by four types of local moves, which are accepted with a probability determined by the Metropolis algorithm::
  \begin{itemize}
    \item Updates of the radial variables $r_{\alpha}$ on a random site. 
    \item Shifts of one of the two random height fields. $h_{\overline{j} \mu \alpha} \rightarrow h_{\overline{j} \mu \alpha} \pm 1$ on a random dual bond
    \item Shifts of the random height fields on two neighboring temporal dual bonds. This is done to assist in thermalization, as it removes intermediate energy barriers required to annihilate certain current loop configurations. The utility of this move is a consequence of the triangular lattice geometry.
    \item On a random dual site, shifts of all the neighboring height fields by $\pm 1$, chosen in a way such that the current loops $p_{j \mu \alpha}$ remain invariant. This effectively constitutes a single-site spin flip of the Ising model that resides on the dual lattice.
  \end{itemize}
We also use several global updates, which we will describe in the subsequent sections. 

To measure the breaking of the $\OO(4)$ symmetry (corresponding to coplanar antiferromagnetism), we define the order parameter $s = \sum_j (r_{j1}^2 - r_{j2}^2)$. This transforms under the adjoint representation of the $\OO(4)$ symmetry - in terms of the original complex spinons $\vec{z}_j$, this is the quantity $\sum_j \vec{z}_j^\dagger \sigma^z \vec{z}_j$ - although it is only this element that remains local under our set of duality transformations described in the previous section. This is a simplification of $\OO(2n)$ models for $n > 1$; for an $\OO(2)$ model in this representation, no such local order parameter exists, and symmetry breaking must be measured through the winding number of non-contractible loops. 
We use the Binder cumulant
\begin{equation}
    U_s \equiv 1 - \frac{\langle s^4 \rangle}{3 \langle s^2 \rangle^2}
\end{equation}
to identify the location of the magnetic phase transition. This quantity approaches unity in the ordered phase (when $\langle s^2 \rangle^2 = \langle s^4 \rangle$) and zero in the disordered phase (when $s$ is a random Gaussian variable with mean zero, $\langle s^4 \rangle = 3 \langle s^2 \rangle^2$). Within the framework of our classical model, the mechanism for the symmetry breaking of $s$ for large $J$ is as follows. As is the case for an $\OO(2)$ model, we have an entropic proliferation of current loops at large $J$. Here, we have two flavors of current loops. A current loop of flavor $\alpha =1$ induces a polarization of the $r_{j}$ variables along that current loop such that it is energetically preferable to have $r_{j1} > r_{j2}$, and analogously for an $\alpha = 2$ current loop. Frustration results from current loops of different flavors intersecting on the same site, and hence it becomes energetically favorable for a single flavor of current loop to coherently proliferate. As we will demonstrate, the complexity of this multi-step mechanism for symmetry breaking leads to large autocorrelation time for $s$ in the ordered phase, as it becomes difficult for $s$ to switch sign once current loops have proliferated. 

Non-magnetic order resulting from gauge field confinement is reflected by lattice symmetry breaking of the gauge-invariant bond frustration, $\varepsilon_{\overline{j}, \mu} \sigma_{\overline{j}, \mu}$. To make connection with prior work studying triangular lattice VBS order~\cite{kaul2015}, we associate an unsatisfied dual bond ($\epsilon_{\overline{j}, \mu} \sigma_{\overline{j}, \mu} = 1$) with the presence of a valence bond $P_{j, \mu} = 1$ on the spatial bond below $(\overline{j}, \mu)$. With this identification, we can calculate the momentum-dependent susceptibility
\begin{equation}
    \chi_{\text{VBS}}(\bm{k}) = \frac{1}{L^3} \sum_{ij} e^{i \bm{k} \cdot (\bm{r}_i - \bm{r}_j)} \langle P_{i, \hat{e}_1} P_{j, \hat{e}_1} \rangle
\end{equation}
where we choose to probe the bond structure on the $\hat{e}_1$ bond. Note that $\bm{k}$ and $\bm{r}_{i, j}$ are two-dimensional vectors specifying only the spatial index. The presence of $\sqrt{12} \times \sqrt{12}$ VBS order is reflected in sharp peaks at the $X$ and $M$ points in the Brillouin zone. We define the quantity $R^M_{\text{VBS}} = 1 - \frac{\chi_{\text{VBS}}(\bm{M} + \bm{a} 2 \pi / L)}{\chi_{\text{VBS}}(\bm{M})}$, which approaches unity in the VBS ordered phase where the Bragg peak becomes infinitely sharp as $L \rightarrow \infty$, but goes to zero in a phase when the height and width of the Bragg peak saturate with system size. The crossing of $R^M_{\text{VBS}}$ for different system sizes is universal and serves as a probe of the location of the VBS phase transition. We also present results for an analogous quantity $R^X_{\text{VBS}}$ which measures the height of the Bragg peak at the X point.
\subsection{Pure \texorpdfstring{$\OO(4)$}{O(4)} model limit}
We first present results for the pure $\OO(4)$ model limit ($K_d = 0$). In this limit, we can compare our numerical results to a classical Monte Carlo simulation of an $\OO(4)$ nonlinear $\sigma$ model. Note that while gauge fluctuations drop out entirely, there is still a non-trivial gauge constraint in our Hilbert space, such that only gauge-invariant observables such as spinon bilinears are non-zero. The critical theory of the phase transition is hence $\OO(4)^*$, which possesses the same critical exponents as the $\OO(4)$ universality class but for which differences can be found in terms of the excitation spectrum for a finite-size system~\cite{whitsitt2016}. This difference is reflected in our dual theory by the topological constraint that the winding number of current loops must be even.

\begin{figure}[htpb]
  \centering
  \includegraphics[width=0.8\textwidth]{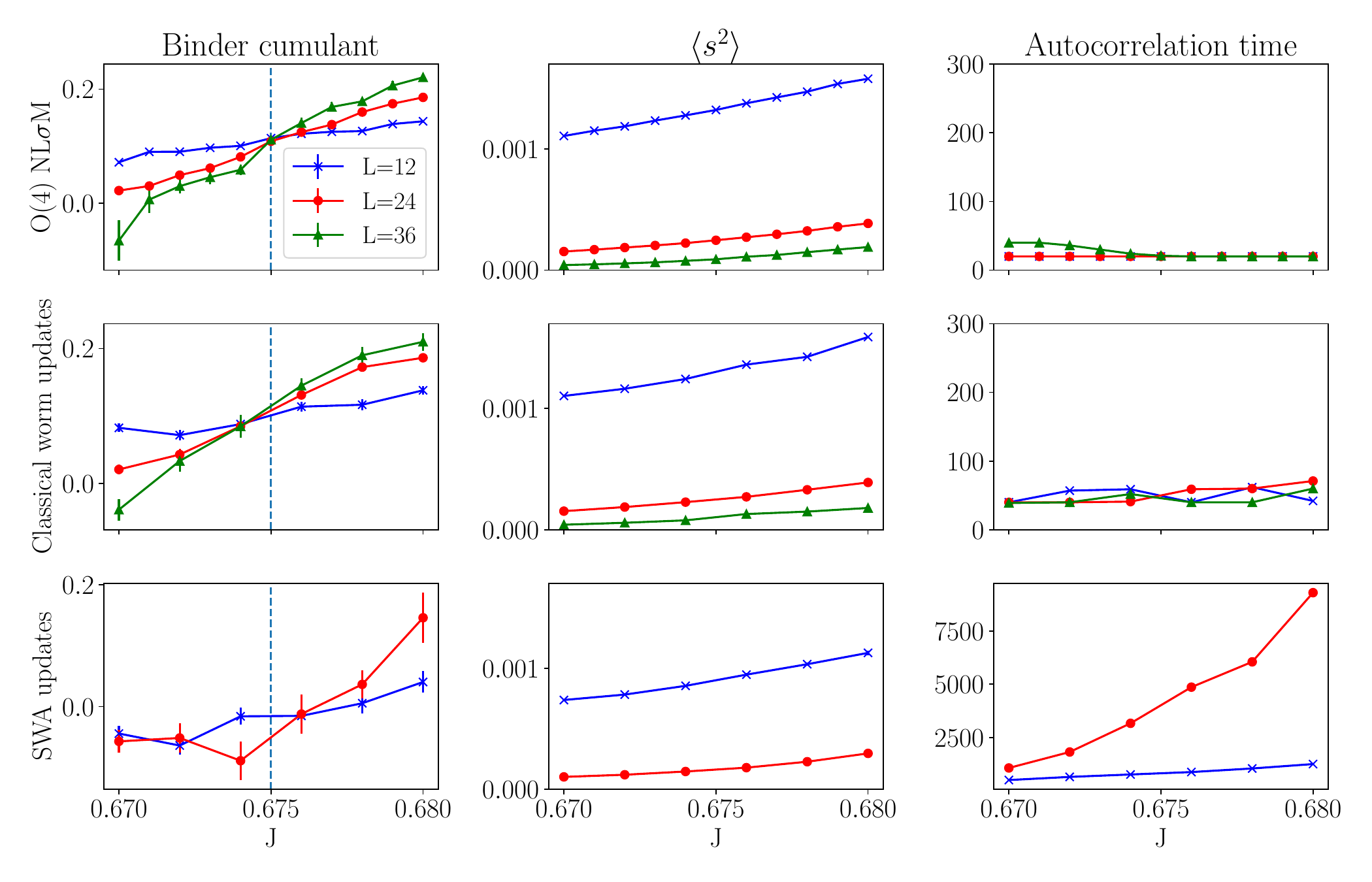}
  \caption{We compare results of numerically simulating the bosonic $\OO(4)^*$ transition while neglecting gauge fluctuations ($K_d = 0$). Direct simulation of an $\OO(4)$ non-linear $\sigma$ model (top row) accurately determines the critical point $J_c \approx 0.675$ while retaining a small autocorrelation time due to global Wolff updates. A dual current loop representation with classical worm updates (middle row) has comparable performance. After restricting to ``surface worm'' updates (bottom row), which are the updates that can be generalized to include gauge fluctuations, we observe a diverging autocorrelation time as we enter the ordered phase and a decrease in accuracy of the Binder cumulant, although an estimate of the location of the critical point can still be inferred.}
  \label{fig:szComparison}
\end{figure}

Recall that for the simpler case of an $\OO(2)$ model, where are dual theory consists only of a single type of integer-valued current loops, classical Monte Carlo simulations which only involve local moves are insufficient for measuring the order parameter for the $\OO(2)$ transition given by a non-zero average winding number of the current loops. These non-contractible loops cannot be obtained from local deformations and must either be generated through a large global update proposal - acceptance of which becomes exponentially unlikely as the system size increases - or through the use of worm algorithms~\cite{prokofev1998, prokofev2001}, where current loops are generated by starting with an ``open'' current string and letting the ends move with suitably-defined probabilities until the two ends meet and form a closed loop. Our generalization to an $\OO(4)$ model naively avoids the need for worm algorithms, as our dual theory retains access to a local order parameter $s$ in addition to the winding number. However, we find that the autocorrelation time of $s$ becomes intractably large, on the order of $10^5$ global sweeps for $L = 24$, near the critical point and into the ordered phase when only local updates are used. This is because, even if one is restricted to the zero winding number sector, large fluctuations in the total current are necessary to induce fluctuations in $s$. The acceptance rate for the creation of local current loops is quite small, on the order of 1-2\% near criticality. Combined with the geometric inefficiency of local current loop updates - creation and annihilation of large current loops require a number of local moves proportional to its area, despite the energy only scaling with the perimeter - leads to this diverging autocorrelation time. Generalizing the classical worm algorithm to account for gauge fluctuations is non-trivial, as one has the constraint that odd current loops must form the boundaries of surfaces of gauge flux. While it is straightforward to apply a classical worm algorithm to a gauge-invariant \textit{pair} of current loops, we find that this, along with an implementation of a ``surface worm algorithm'' (SWA) proposed in~\cite{delgadomercado2013} and summarized in Appendix~\ref{app:worm}, are insufficient for reducing the autocorrelation time to a tractable magnitude. This is because the propagation of a \textit{pair} of current loops is much more energetically costly than a single current. An appropriate worm algorithm in the limit of weak gauge fluctuations would be to propagate a worm as normal, calculate the energy cost of an enclosed surface (ideally the minimal surface) once the worm has terminated, and accept the worm with a probability determined by the energy cost of the surface. However, implementing this algorithm is challenging as it requires an efficient way of finding a candidate surface once the worm has been grown. We leave further development of this approach to future work. 

In Fig.~\ref{fig:szComparison}, we present the Binder cumulant, the order parameter $\langle s^2 \rangle$, and the autocorrelation time from direct simulations of an $\OO(4)$ NL$\sigma$M as compared to simulations in the dual theory, using either a classical worm algorithm (only appropriate in the $K_d=0$ limit) or a SWA. While our simulation with a SWA is able to identify the location of the critical point $J_c \approx 0.675$ with reasonable accuracy, the large autocorrelation time prevents us from both going to larger system sizes and obtaining high precision results for the Binder cumulant.


\subsection{Pure gauge theory limit}
We also consider the pure gauge theory limit ($J \rightarrow 0$), which is dual to an Ising model with spatial frustration. In this limit, all current loops cost infinite energy, and our state space is restricted to configurations with $p_{j \mu \alpha} = 0$. Within this space, we have a single $\mathbb{Z}_2$ degree of freedom residing on each dual site $(\overline{j})$, the flipping of which at site $\overline{k}$ corresponds to the shift $h_{\overline{j} \mu \alpha} \rightarrow h_{\overline{j} \mu \alpha} + \Delta_{\mu} f_{\overline{j} \alpha}$, with $f_{\overline{j} \alpha} = \delta_{\overline{j} \overline{k}}$ (performed on a random choice of $\alpha$). Since the action is only sensitive to the parity of $\sum_{\alpha} h_{\overline{j} \mu \alpha}$, this is effectively a $\mathbb{Z}_2$ degree of freedom. Our model then reduces down to a frustrated Ising model defined on the dual (hexagonal) lattice, the simulation of which is carried out using local spin flips and a global cluster update similar to the one described in~\cite{moessner2001}. We define our cluster update explicitly in Appendix~\ref{app:ising}. Note that in contrast to the pure $\OO(4)$ model limit where we provided a numerical comparison with results from an $\OO(4)$ NL$\sigma$M simulation, we do not provide an analogous comparison here as the degrees of freedom in our model explicitly corresponds to that of a frustrated Ising model.

\begin{figure}[htpb]
    \includegraphics[width=0.45\textwidth]{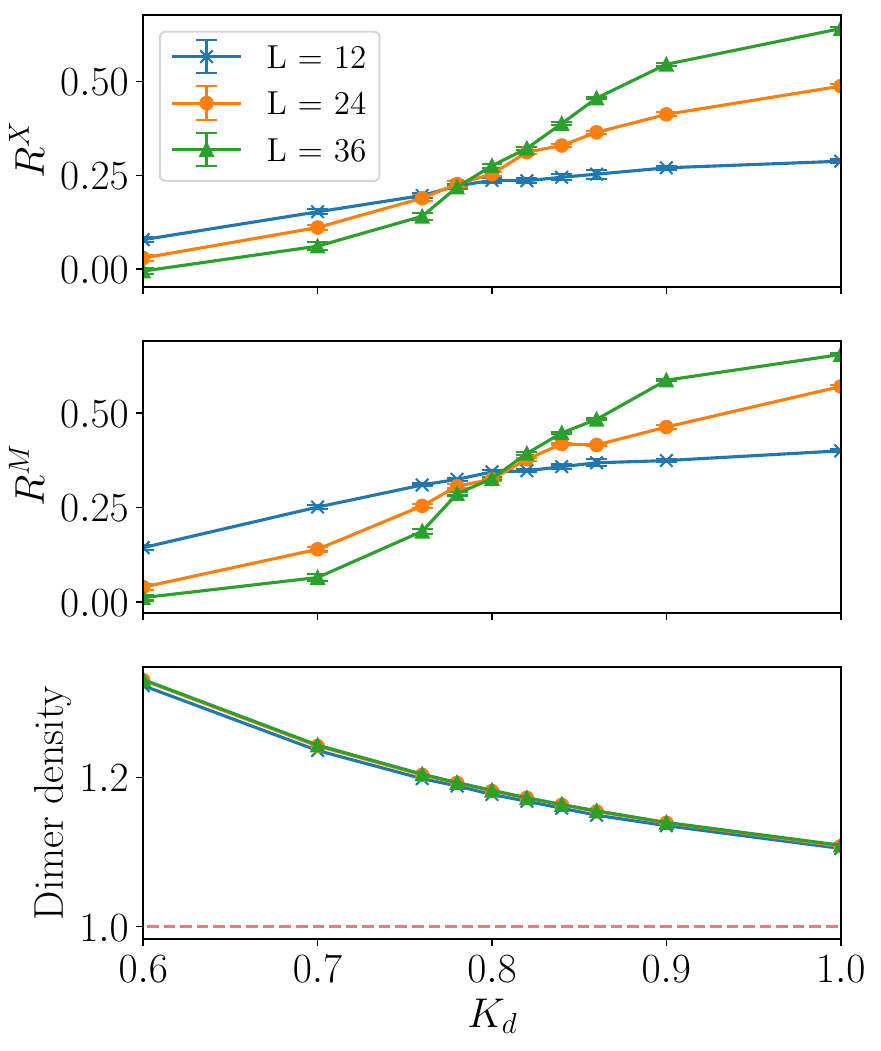}
    \includegraphics[width=0.35\textwidth]{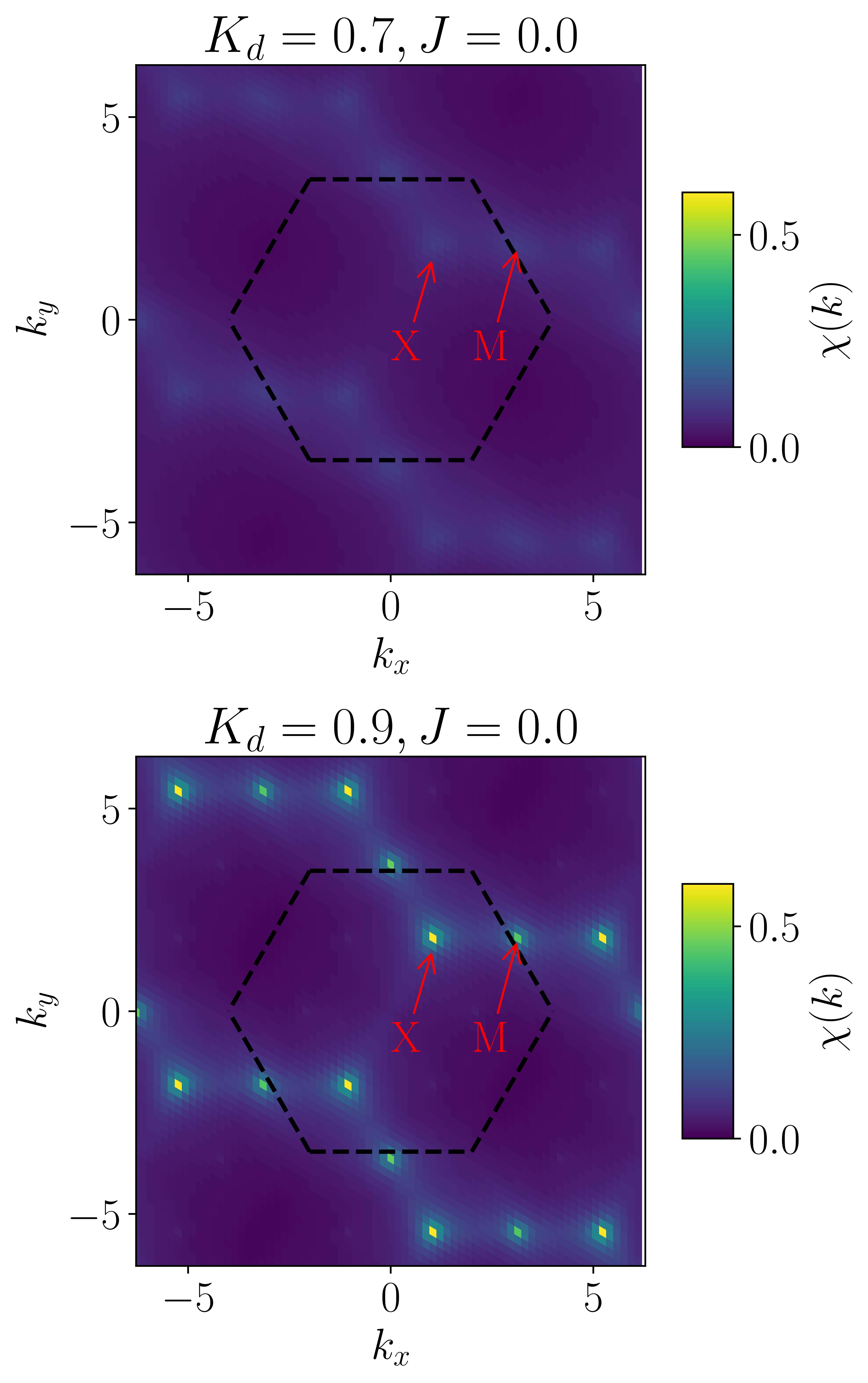}
    \caption{We present results of simulating an odd $\mathbb{Z}_2$ gauge theory on a triangular lattice, which displays a transition from a deconfined phase for $K_d < K_d^c$ and a VBS phase with $\sqrt{12} \times \sqrt{12}$ order for $K_d > K_d^c$, with $K_c \approx 0.82$. The equal-time dimer correlation function, plotted here for $L = 36$, displays sharp peaks at the $M$ and $X$ points in the ordered phase.}
    \label{fig:pureGauge}
\end{figure}

The phase diagram of the fully frustrated quantum Ising model on the hexagonal lattice has been analyzed theoretically~\cite{gregoire2008, coletta2011}, although no numerical studies have been conducted to our knowledge aside from results at a single point in the ordered phase in~\cite{moessner2001}. In the limit of small transverse field, this is equivalent to a triangular lattice quantum dimer model with interaction strength $V$ and hopping $t$ at the point $V/t=0$. These models have been studied much more extensively~\cite{ralko2005, ralko2006} and the existence of a $\sqrt{12} \times \sqrt{12}$ phase has been well-established. We note that path integral Monte Carlo simulations of this two dimensional quantum Ising model in terms of a three dimensional classical model pose much more challenges due to the inherent discretization errors not present in dimer model simulations. In particular, we find that an isotropic scaling of the spatial and temporal couplings $K_d^s\,, K_d^\tau$ leads to very weak ordering. This is because for a system size finite in the temporal direction, increasing $K^\tau_d$ simultaneously increases the antiferromagnetic interaction strength as well as the effective temperature of the quantum model. In order to obtain a clear signature of the phase transition, we parameterize the couplings as
\begin{equation}
  \begin{aligned}
  K_d^s &= K_d \\
  K_d^\tau &= \text{min}\left\{K_d\,, \frac{\ln 2}{2}\right\}
  \end{aligned}
\end{equation}
in order to prevent $K_d^\tau$ from getting too large; the particular value of $\frac{\ln 2}{2}$ is chosen such that the quantum temperature (in units of the transverse field) is equal to the inverse length of the system size in the temporal direction.

We present numerical results in Fig.~\ref{fig:pureGauge}. We identify a transition into a $\sqrt{12} \times \sqrt{12}$ ordered phase at $K_d \approx 0.82$, in surprisingly good agreement with semiclassical analyses~\cite{gregoire2008} of the quantum Ising model which predict $K_d = \frac{2}{\sqrt{6}} \approx 0.816$. We also plot the ``dimer density,'' defined as the number of unsatisfied spatial bonds per site. This should approach the minimum value of one and hence reduce to the $V/t = 0$ quantum dimer model in the $K_d \rightarrow \infty$ limit, which agrees with our numerical results.
\subsection{Antiferromagnet to VBS transition}
\begin{figure}[htpb]
  \centering
  \includegraphics[width=0.55\textwidth]{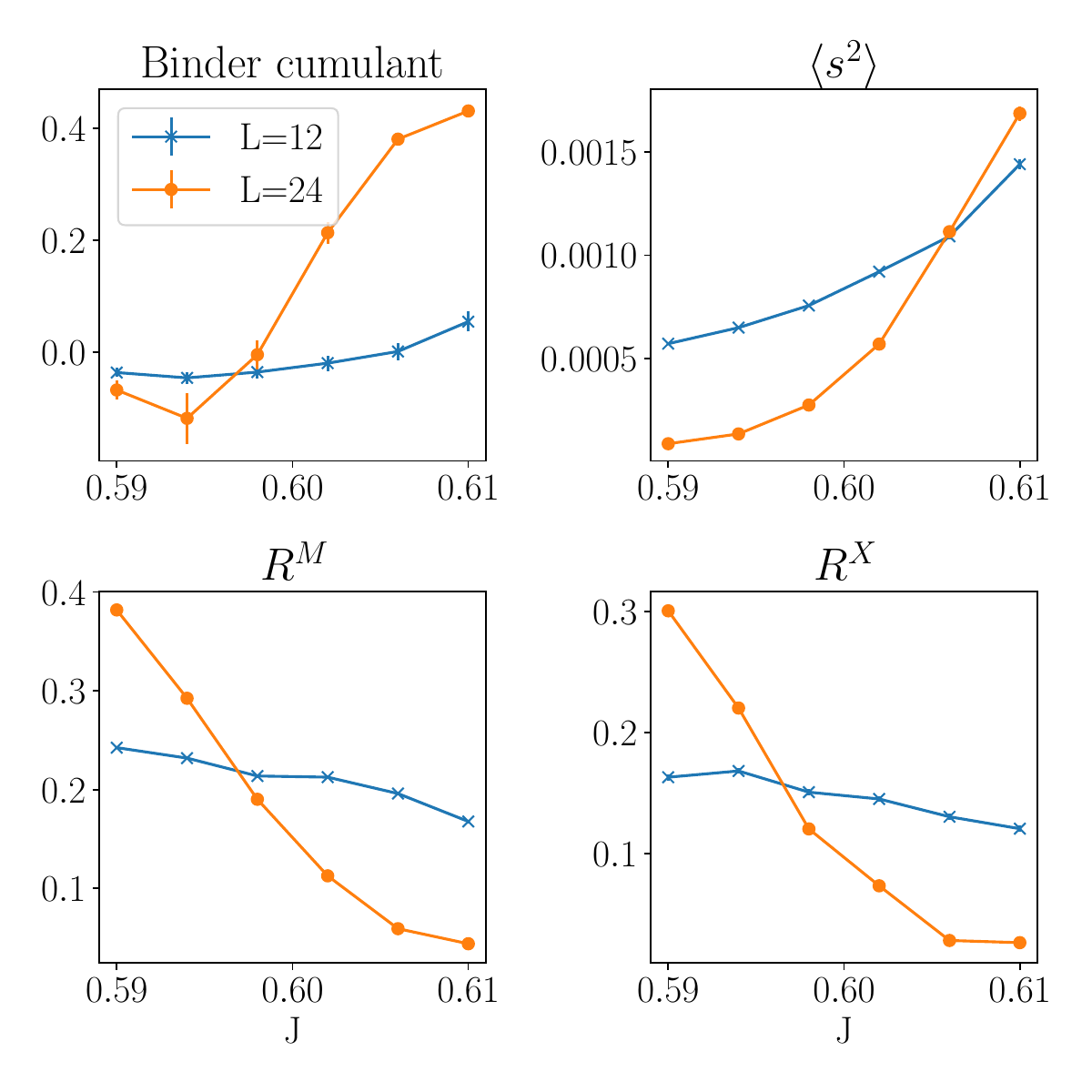}
  \includegraphics[width=0.35\textwidth]{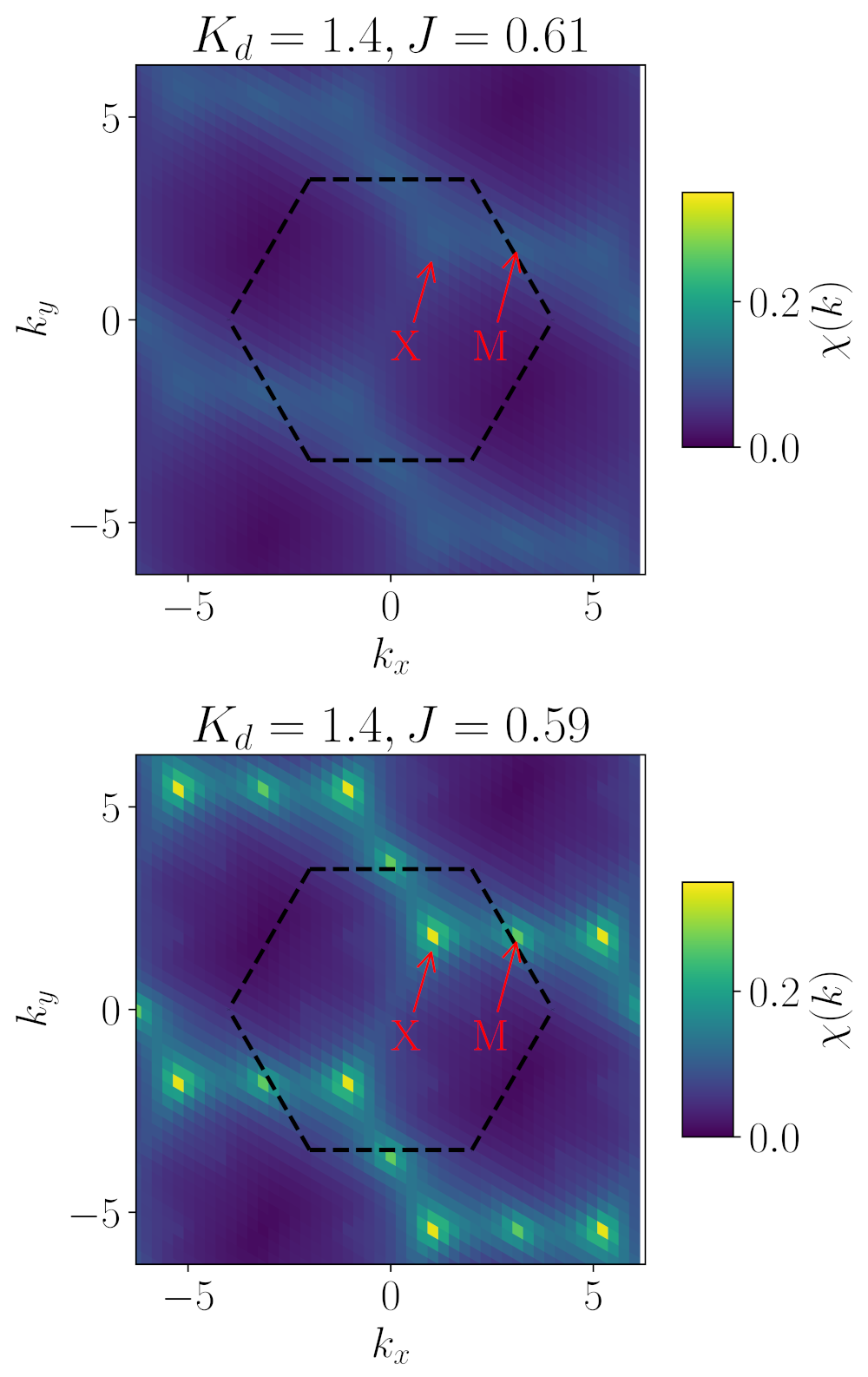}
  \caption{We present measurements of the antiferromagnetic order parameter $s$ and its Binder cumulant, along with measures of the VBS order at the $X$ and $M$ points in the Brioullin zone, at fixed gauge coupling $K_d = 1.4$ as a function of $J$. The crossing of the Binder cumulant at $J_c \approx 0.6$ closely coincides with the loss of VBS order, suggesting a direct transition between the two phases. On the right, we plot the equal-time dimer correlation function $\chi(k)$, demonstrating the loss of order at the $M$ and $X$ points for $J > 0.6$.}
  \label{fig:dqcp}
\end{figure}

\begin{figure}[htpb]
  \centering
  \includegraphics[width=0.55\textwidth]{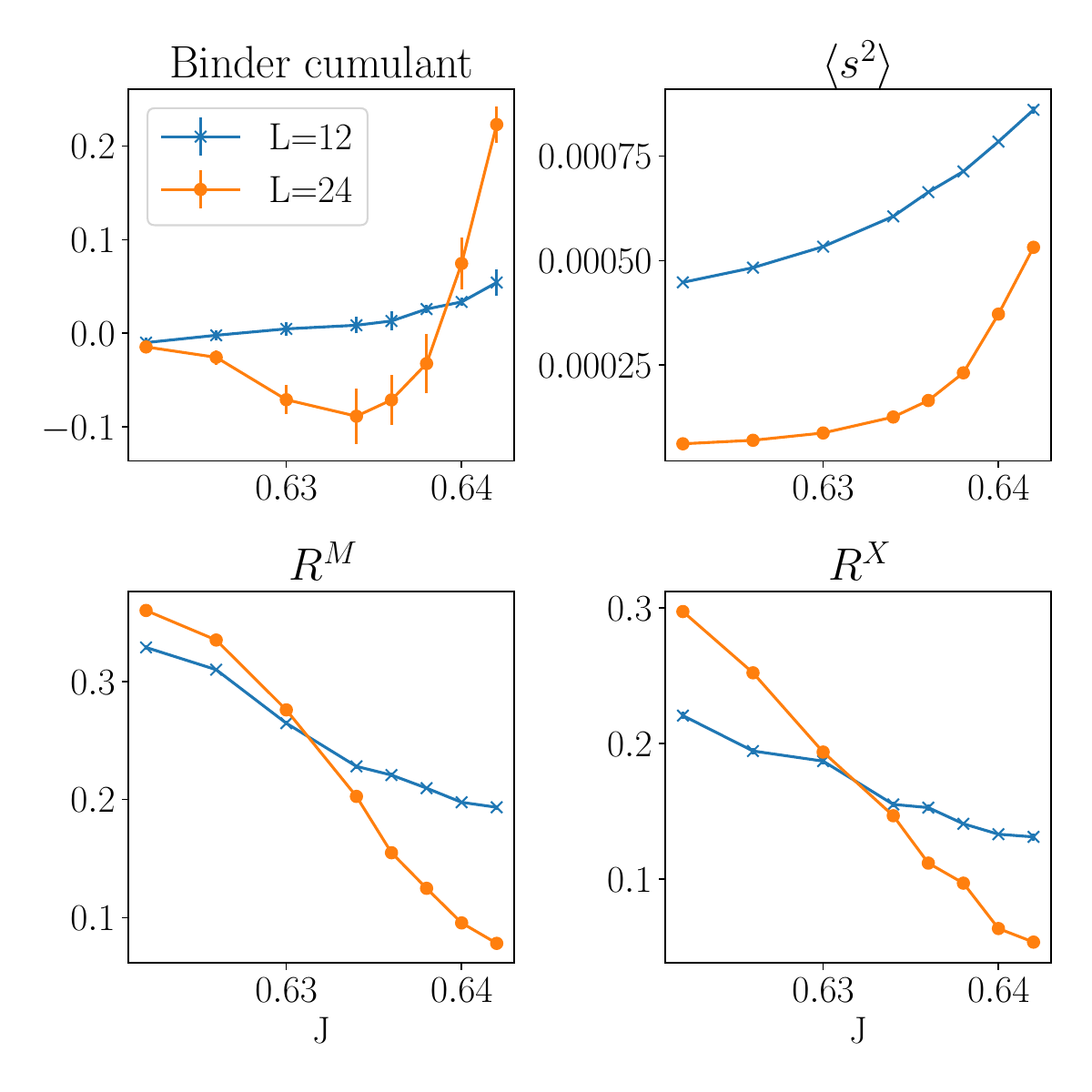}
  \includegraphics[width=0.35\textwidth]{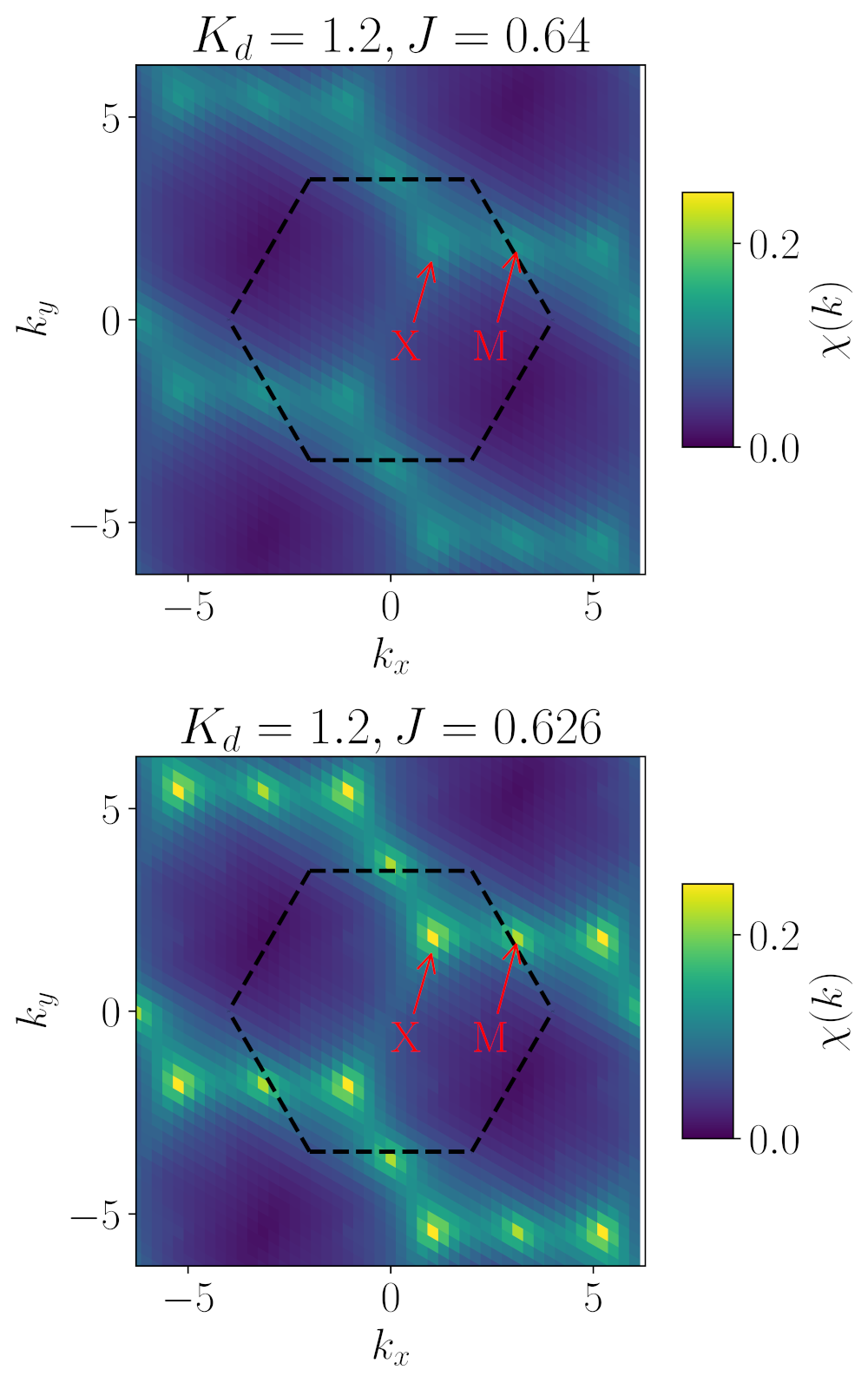}
  \caption{We plot the same data as in Fig.~\ref{fig:dqcp}, but for weaker gauge fluctuations, $K_d = 1.2$. We note a larger gap between the vanishing of the VBS order and the appearance of antiferromagnetic order, suggesting that the $\mathbb{Z}_2$ spin liquid phase might persist in this region; however, larger system sizes may yield a direct transition.}
  \label{fig:dqcp2}
\end{figure}

Having established the validity of our sign-problem-free model in the limit where either the spinons or visons are static, we now analyze the transition from VBS to antiferromagnetic order, obtained by starting in the VBS phase and increasing $J$ until the proliferation of current loops destroys the effective dual Ising model.This is plotted across two slices, $K_d = 1.4$ in Fig.~\ref{fig:dqcp} and $K_d = 1.2$ in Fig.~\ref{fig:dqcp2}. We find that the loss of VBS order closely coincides with growing antiferromagnetic order, indicating a direct transition. However, we emphasize that the difficulties present in establishing the antiferromagnetic transition for $K_d = 0$ also persist for $K_d > 0$ - the autocorrelation time for the antiferromagnetic order parameter $s$ quickly diverges as we approach the magnetically ordered phase, which prohibits us from reaching larger system sizes. As a consequence, while we observe some signatures of a first-order transition - including the Binder cumulant dipping below zero near the phase transition, along with with a sharper upturn in $s^2$ which may evolve into a discontinuous jump for larger system sizes - our numerical results currently cannot definitely establish the nature of this transition.

For even larger $K_d$, we expect for current loops to proliferate at smaller values of $J$, as the $\order{K_d}$ energy penalty incurred by unsatisfied bonds in the dual Ising model can be alleviated through the presence of current loops. In particular, as $K_d \rightarrow \infty$, we expect the antiferromagnetic phase to persist for any non-zero value of $J$. This behavior was found in an analogous simulation of an $\OO(2)$ model on a square lattice~\cite{park2002}, where the current loop proliferation at $K_d \rightarrow \infty$ is a consequence of the superfluid instability of the Bose-Hubbard model at half-integer filling to arbitrarily weak hopping. We intend to investigate this limit further in future work, as specialized updates become necessary for large $K_d$.
\section{Outlook}
In this work, we present an effective model of quantum antiferromagnetism on the triangular lattice and demonstrate that it can be mapped to a classical sign-problem-free partition function. This extends known duality mappings~\cite{park2002, endres2007, delgadomercado2013} and yields sign-problem-free models for a broad class of systems, of which our effective model is only one example of. Additional models are of interest for future research. In particular, our effective model may be defined on a kagom\'e lattice, where a similar effective description of $\OO(4)$-symmetric bosons coupled to an odd $\mathbb{Z}_2$ gauge field exists~\cite{sachdev1992}. Numerical studies of extended Heisenberg models on the kagom\'e lattice~\cite{wietek2020} have given evidence for a ``diamond'' VBS order - it would be fruitful to study whether such a VBS pattern can naturally emerge from $\mathbb{Z}_2$ gauge field confinement with a background Berry phase. As our mapping also applies to bosons at \textit{generic} fillings - not just a fixed background of one boson per site - it is also of interest to study the effects that a $\mathbb{Z}_2$ gauge field has on the phase diagram of the Bose-Hubbard model. Recent experimental proposals~\cite{homeier2023} for realizing such models in quantum simulators make this question of timely importance.

Additionally, there is much room for algorithmic improvements in Monte Carlo simulations of these models. With the current system sizes accessible to us, we are able to establish the existence of spin liquid, VBS, and antiferromagnetic phases. However, the nature of the VBS to antiferromagnet transition is unclear; our results support a first-order transition, but the small number of system sizes accessible to us along with a diverging autocorrelation time leaves open the possibility of a continuous transition. Broadly-applicable techniques such as parallel tempering and reweighting may somewhat improve numerical results, but our classical model presents a fundamental complexity arising from the competition between multiple types of degrees of freedom. We have implemented several global updates to improve sampling, but we expect that more sophisticated sampling methods would have more success in clearly resolving the putative DQCP. It would also be useful to consider formulations in terms of alternate degrees of freedom. One of the difficulties that prevent robust global updates is the geometric complexity in the model, where degrees of freedom effectively live on sites, bonds, dual sites, and dual bonds. This is in contrast to J-current formulations used for simulating $\text{NCCP}^1$ models~\cite{troyer2008}, where all the degrees of freedom live on bonds and good worm algorithms exist.   
We expect that a continuous-time generalization of this classical model would also improve performance as well as remove Trotterization errors and connect more directly to the underlying quantum model.

\subsection*{Acknowledgements}
We thank Ribhu Kaul, Cenke Xu, Peter Lunts, Lukas Weber, and Vera Ko for useful discussions.
This research was supported by the U.S. National Science Foundation grant No. DMR-2245246 and by the Simons Collaboration on Ultra-Quantum Matter which is a grant from the Simons Foundation (651440, S.S.). 

\appendix
\section{Derivation of effective model}%
\label{sec:derivation}

For completeness, we summarize the derivation of our effective model, first calculated in~\cite{sachdev1992}. Our starting point is the spin-$1 / 2$ Heisenberg antiferromagnet on the triangular lattice,
\begin{equation}
  \begin{aligned}
    H = \sum_{i j } J_{i j} \vec{S}_i \cdot \vec{S}_j\,,
    \label{eq:antiferroApp}
  \end{aligned}
\end{equation}
with $J_{i j}$ short-ranged antiferromagnetic interactions. In our derivation, we will take $J_{ij} = J$ on nearest-neighbor sites, and $0$ otherwise. We use a Schwinger boson representation, where the $(2S+1)$ states of a spin-$S$ representation of $\SU(2)$ can be represented in terms of bosonic operators $s_{i \uparrow}$, $s_{i \downarrow}$,
\begin{equation}
  \begin{aligned}
    \ket{S\,, m} = \frac{1}{\sqrt{(S+m)!(S-m)!} } \left( s^\dagger_{i\uparrow} \right)^{S+m} \left( s^\dagger_{i \downarrow} \right)^{S-m} \ket{0}
  \end{aligned}
\end{equation}
where $m= -S\,, \ldots \,, S$ is the $z$ component of the spin, and the vacuum $\ket{0}$ contains no Schwinger bosons. Our physical Hilbert space is obtained by the restriction
\begin{equation}
  \begin{aligned}
  s_{i \alpha}^\dagger s_{i}^{\alpha} = 2 S \equiv n_s\,.
  \end{aligned}
\end{equation}
The $n_s \rightarrow \infty$ limit is classical and results in non-collinear antiferromagnetic order. In order to retain quantum fluctuations, we additionally generalize the $\SU(2)$ symmetry to $\USP(2M)$ and take the limit $n_s\,, M \rightarrow \infty$ with $\kappa = n_s / M$ fixed. The generalization to $\USP(2M)$ rather than $\SU(2M)$ is chosen to ensure the existence of a spin singlet state, given by
\begin{equation}
  \begin{aligned}
    \mathcal{J}^{\alpha \beta} s_{i\alpha}^\dagger s_{j \alpha}^\dagger \ket{0}\,,
  \end{aligned}
\end{equation}
with $\mathcal{J}^{\alpha \beta}$ a $2M \times 2M$ matrix,
\begin{equation}
  \begin{aligned}
    J = \begin{pmatrix} & 1 & & & & & \\ - 1 & & & & & \\ & & & 1 & & \\ & & -1 & & & \\ & & & & & \ldots \\ & & & & \ldots &  \end{pmatrix} 
  \end{aligned}
\end{equation}
and the $\USP(2M)$ group defined by the set of unitary matrices $U$ that satisfy $U^T \mathcal{J} U = \mathcal{J}$.

Writing our Hamiltonian in Eq.~\ref{eq:antiferroApp} in terms of Schwinger bosons,
\begin{equation}
  \begin{aligned}
  H = - \sum_{i > j} \frac{J_{ij}}{2M} \left( \mathcal{J}^{\alpha\beta} s_{i \alpha}^\dagger s_{j \beta}^\dagger \right)  \left( \mathcal{J}_{\gamma\delta} s_{i}^\gamma s_j^\delta \right) 
  \end{aligned}
\end{equation}
moving to a path integral representation,
and performing a Hubbard-Stratonovich transformation to decouple the four-boson term, we obtain
\begin{equation}
  \begin{aligned}
    Z &= \int \calD \mathcal{Q} \calD s \calD \lambda \exp \left( - \int_0^\beta \mathcal{L} \dd{\tau} \right) \,,
    \\
  \mathcal{L} &= \sum_i \left[ s_{i\alpha}^\dagger \left( \dv{\tau} + i \lambda_i  \right) s_{i}^{\alpha} - i \lambda_i n_s \right] 
    \\
    &+ \sum_{\langle i j \rangle} \left[ M \frac{J_{ij} \abs{\mathcal{Q}_{ij}}^2}{2} - \frac{J_{ij}\mathcal{Q}^*_{ij}}{2} \mathcal{J}^{\alpha\beta} s_{i \alpha} s_{j \beta} + \text{h.c} \right] \,.
    \label{eq:largeMLagrangian}
  \end{aligned}
\end{equation}
This Lagrangian has $\UU(1)$ gauge invariance, under which
\begin{equation}
  \begin{aligned}
    s_{i\alpha}^\dagger &\rightarrow s_{i\alpha}^\dagger \exp(i \rho_i(\tau))\,, \\
    \mathcal{Q}_{ij} &\rightarrow \mathcal{Q}_{ij} \exp \left( i \rho_i(\tau) - i \rho_j(\tau) \right) \,, \\
    \lambda_i &\rightarrow \lambda_i + \pdv{\rho_i}{\tau}(\tau) \,.
  \end{aligned}
\end{equation}
The saddle-point solutions $\overline{\mathcal{Q}}$, $\overline{\lambda}$ of Eq.~\ref{eq:largeMLagrangian} have been obtained previously~\cite{sachdev1992}. The saddle-point values $\overline{Q}$ satisfy
\begin{equation}
  \begin{aligned}
  \overline{Q}_{ij} = \langle \mathcal{J}_{\alpha\beta} s_i^\alpha s_j^\beta \rangle \,,
  \end{aligned}
\end{equation}
which imply anti-symmetry under exchange of $i$ and $j$. These saddle-point solutions can be chosen to satisfy $\mathcal{Q}_{i\,, i + \hat{e}_p} = \overline{\mathcal{Q}}$, $i \lambda_i = \overline{\lambda}$, where the unit vectors $\hat{e}_p$
\begin{equation}
  \begin{aligned}
  \hat{e}_1 &= \left( 1 / 2\,, \sqrt{3} /2 \right)  \\
  \\
  \hat{e}_2 &= \left(  1 / 2\,, -\sqrt{3} /2 \right)  \\
  \\
  \hat{e}_3 &= \left( -1 \,, 0  \right) \\
  \end{aligned}
\end{equation}
point between nearest neighbor sites. The anti-symmetry under exchange of $i$ and $j$ implies that the mean-field solution for $\mathcal{Q}$ will break reflection symmetry; however, reflection symmetry can be restored by a gauge transformation. Of note are non-translationally-invariant saddle-point solutions for $\mathcal{Q}$~\cite{huh2011}, whose solutions $\mathcal{Q}_{ij}^v$ relative to the translationally-invariant saddle-point correspond to a localized defect, along with a ``branch cut'' extended outwards from the core, where $\text{sgn}(\mathcal{Q}_{ij}^v ) = - \text{sgn}(\overline{\mathcal{Q}}_{ij})$. These saddle-point solutions are identified with gapped vison excitations in the corresponding $\mathbb{Z}_2$ spin liquid, whose exchange statistics with the Schwinger bosons are mutual semions.

Taking these saddle-point solutions, the hopping term $\mathcal{J}^{\alpha\beta} s_{i \alpha} s_{j \beta}$ can be diagonalized, leading to a continuum Lagrangian
\begin{equation}
  \begin{aligned}
    \mathcal{L} &= x_\alpha^* \pdv{x_\alpha}{\tau} + y_\alpha^* \pdv{z_\alpha}{\tau} + z_\alpha^* \pdv{y_\alpha}{\tau} + \left( \overline{\lambda} - 3 \sqrt{3}  J \overline{\mathcal{Q}}/  2 \right) \abs{z_\alpha}^2
    \\
    &+ \left( \overline{\lambda} + 3 \sqrt{3}  J \overline{\mathcal{Q}} / 2 \right) \abs{y_\alpha}^2 + \overline{\lambda} \abs{x_\alpha}^2 + \frac{3 J \overline{\mathcal{Q}}}{2} \left( \abs{\partial_x z_\alpha}^2 + \abs{\partial_y z_\alpha}^2 \right) + \ldots
  \end{aligned}
\end{equation}
where we have written our bosonic spinons $s_{i \alpha}$ in terms of three variables $x_\alpha\,, y_\alpha\,,z_\alpha$, related by a unitary
transformation to the three bosonic spinons on the three site of each unit cell. The bosonic spinon $z_\alpha$ has the lowest mass,
and hence the transition between the theory with antiferromagnetic long-range order ($\langle \vec{S_i} \rangle \neq 0$) and 
the quantum-disordered phase is driven by the condensation of $z_\alpha$. The other spinon fields can be integrated out, yielding
the effective Lagrangian
\begin{equation}
  \begin{aligned}
    \mathcal{L} &= \frac{1}{\overline{\lambda} + 3 \sqrt{3} J \overline{\mathcal{Q}}/2} \abs{\partial_\tau z_\alpha}^2 + \frac{3 J \overline{\mathcal{Q}} \sqrt{3} }{8} \left( \abs{\partial_x z_\alpha}^2 + \abs{\partial_y z_\alpha}^2 \right) 
    \\
    &+ \left( \overline{\lambda} - 3 \sqrt{3} J \overline{\mathcal{Q}} /2 \right) \abs{z_\alpha}^2+ \ldots\,.
  \end{aligned}
\end{equation}

Provided the visons remain gapped, this theory describes a deconfined critical point separating a state with long-range antiferromagnetic order to an odd $\mathbb{Z}_2$ spin liquid. One may additionally consider a possible vison condensation, where the vison Berry phase will lead to valence bond solid ordering. These two transitions can be captured in the partition function
\begin{equation}
  \begin{aligned}
    \mathcal{Z} &= \sum_{s_{j, j + \hat{\mu}} = \pm 1} \prod_j \int \dd{z}_{j \alpha}\delta \left( \sum_\alpha \abs{z_{j \alpha}^2} - 1 \right) \left[ \prod_j s_{j, j + \tau} \right]^{2S} \exp\left( - H[z_\alpha, s] \right)  \\
    H[z_\alpha, s] &= - \frac{J}{2} \sum_{ \langle j , \mu \rangle} s_{j, j + \hat{\mu}} \left( z_{j,  \alpha}^* z_{j + \hat{\mu},  \alpha} + \text{ c.c} \right) - K \sum_{\triangle \square} \prod_{\triangle \square} s_{j, j + \hat{\mu}} \,,
  \end{aligned}
\end{equation}
where we have introduced the $\mathbb{Z}_2$ gauge field $s_{ij}$ defined on the links of our lattice. The model is defined on a three-dimensional stacked triangular lattice, where we have discretized our temporal direction. For large $J$, fluctuations of $z$ are suppressed and we recover N\'eel order. For small $K$, vison excitations proliferate and we obtain either a trivial phase (integer $S$) or valence bond solid order (half-integer $S$). Importantly, we must include a Berry phase term $\prod_j s_{j, j + \tau}$, which is non-trivial for half-integer spin. In particular, this Berry phase is an obstacle for using classical Monte Carlo methods to evaluate the partition function for half-integer spin, as the sign of each term in the partition function may be either positive or negative and hence prevents evaulation via probabilistic sampling. One of the results of this work is to derive a sign-free representation of Eq.~\ref{eq:bosonZ} amenable to Monte Carlo studies.

\section{Details of numerical simulations}
\label{app:numerics}
Here, we provide additional information regarding Monte Carlo simulations of our effective model. A single Monte Carlo simulation consists of $10^6$ sweeps, where a single sweep consists of $L^3$ of each of the local and cluster updates described in the main text. The first 50\% of sweeps are used to thermalize the system. All measurements are averaged over $100$ runs with different random seeds. We use the Xoshiro256+ algorithm for generating random numbers. In order to reduce the computational bottleneck arising from repeated evaluations of the modified Bessel function $I_p(x)$ present in our partition function, we pre-compute a lookup table of size $10 \times 10^4$ for integer values of $0 \leq p < 10$ and a discretized grid of size $10^4$ of $x$ values between $0$ and the maximum possible value of $x$, $J / 2$. With this, the majority of the computation time is spent computing geometric information, such as finding nearest neighbor sites or the bonds surrounding a plaquette. A large amount of geometric data is relevant for our simulations as we work with sites, bonds, dual sites, and dual bonds; as a result, pre-computing all the required geometric data and storing it in memory leads to a large number of cache misses and is ultimately slower than computing the information each time.

\section{Surface worm algorithm}
Details of the surface worm algorithm (SWA), first discussed in~\cite{delgadomercado2013}, are presented here.

\begin{figure}[htpb]
  \centering
  \includegraphics[width=0.8\textwidth]{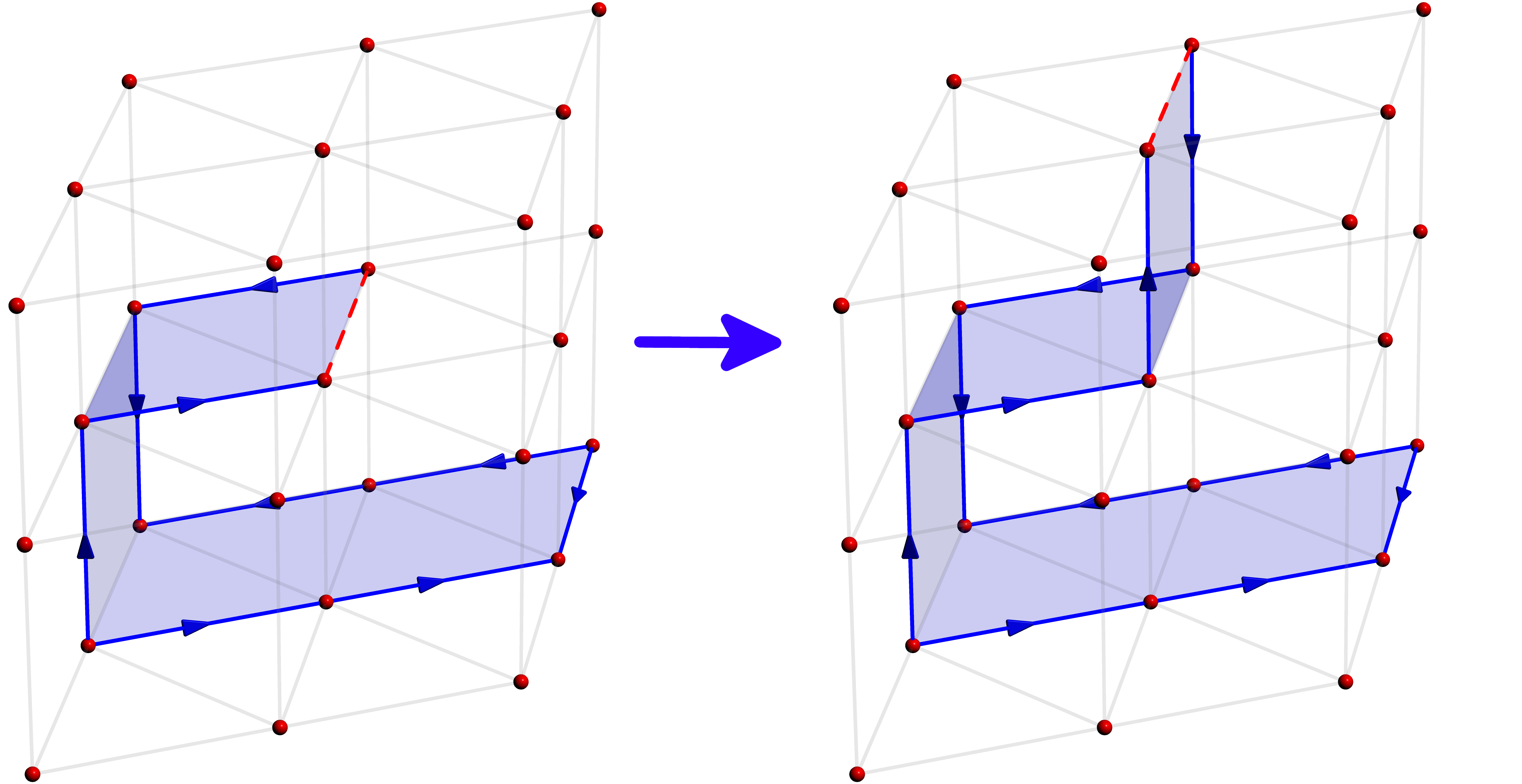}
  \caption{We illustrate the growing of a ``surface worm'' using the SWA. At each step, the worm can either attempt to grow in a random direction or attempt to close the loop. Each move is accepted probabilistically according to the Metropolis algorithm. Note that this worm can move in all directions, not just the two-dimensional plane illustrated.}
  \label{fig:SWA}
\end{figure}

The idea behind the SWA, as with worm algorithms more generally, is to generate large moves via probabilistically moving through unphysical configurations such that the final physical configuration obeys detailed balance with respect to the original one. Traditional worm algorithms are applied to systems where physical configurations correspond to some form of closed loops, and the unphysical configurations that the algorithm moves through are ones with an open loop. For gauge-Higgs models, this simple application is not appropriate - the closed loops correspond to bosonic worldlines, and the presence of a gauge field means that worldlines of charged operators must form the boundary of a surface of gauge flux. Rather than growing a single current, the SWA grows a ``ligament'' corresponding to two parallel current loops bounded by flux. This process is illustrated in Fig~\ref{fig:SWA}.

There is a geometric subtlety in implementing the SWA here, which arises from the non-cubic lattice structure. For a cubic lattice, all possible moves from a given bond are chosen with equal probability. For a stacked triangular lattice, there is a difference depending on whether the constraint-violating bond is a spatial and temporal bond - the former has 10 possible moves, whereas the latter has 18. In order to maintain detailed balance, the probability of picking a temporal bond while on a spatial one must equal the probability of picking a spatial bond while on a temporal one. To enforce this constraint, all 18 moves from a temporal bond are chosen with equal probability, and the moves from a spatial bond are chosen in a skewed manner such that the probability of moving to any of the four neighboring temporal bonds is $ 4 \times \frac{1}{18}$.
\label{app:worm}

\section{Cluster algorithm for dual Ising model}
\label{app:ising}
Here, we provide more details on the cluster updates we use for the dual Ising degrees of freedom. This style of updates was described in~\cite{moessner2001}. The update we use is a variant on the Wolff cluster update~\cite{wolff1989}, a well-known cluster algorithm for efficiently generating global moves in Ising models. However, this algorithm becomes inefficient in the presence of frustration. As our dual Ising model only has frustration in the spatial bonds, we adapt the algorithm such that cluster are only grown along the frustration-free temporal bonds - a standard Wolff algorithm may still in principle be used and will lead to comparable convergence times when measured in terms of Monte Carlo steps, but will be significantly more computationally demanding than this more targeted cluster update.

We apply this algorithm to both single dual sites and pairs of dual sites - the latter is necessary as movement between low-energy configurations is accomplished by flipping neighboring pairs of spins. For a single site update, we pick a dual site $\overline{j}$ at random and calculate the energy $\Delta E_s$ incurred from the spatial bonds after flipping the Ising degree of freedom on that site. This site is then added to our cluster. We grow the cluster in the temporal direction, where growing a cluster in the $\pm \tau$ direction is accepted with probability $p = \text{min}\{0, 1 - e^{-2K_d^\tau \sigma_{j} \sigma_{j \pm \tau}}\}$. The energy from the spatial bonds of these spins are added to $\Delta E_s$. Once the cluster has finished growing, the entire cluster is flipped with probability $\text{min}\{0, e^{-\Delta E_s}\}$. This illustrates the necessity for keeping the temporal coupling $K_d^\tau$ relatively small, as a sufficiently large $K_d^\tau$ will lead to clusters spanning the entire temporal direction and our model effectively reduces to that of a classical 2D Ising model.

For performing this update on a pair of neighboring dual sites $\overline{j}\,, \overline{k}$, we proceed in an identical fashion, growing of a cluster in the $\pm \tau$ direction with probability $p = \text{min}\{0, 1 - e^{-2K_d^\tau(\sigma_{j} \sigma_{j \pm \tau} + \sigma_k \sigma_{k \pm \tau})}\}$. The inclusion of these moves are important as $\Delta E_s$ will generally be much smaller for these moves.
\bibliography{O4}

\end{document}